\documentclass[12pt]{article}
\usepackage{graphicx}
\usepackage{amssymb}
\usepackage{bm}
\usepackage{amsfonts}
\usepackage{amsmath}
\usepackage{cases}
\usepackage{hyperref}

\newcommand{\beq}{\begin{equation}}
\newcommand{\eeq}{\end{equation}}
\newcommand{\beqa}{\begin{eqnarray}}
\newcommand{\eeqa}{\end{eqnarray}}
\newcommand{\w}{\wedge}

\newcommand{\ts}{\textstyle}

\newcommand{\h}[1]{\hat{#1}}
\newcommand{\ou}[3]{\underset{#3}{\overset{#1}{#2}}}
\newcommand{\ua}{\uparrow}
\newcommand{\da}{\downarrow}
\newcommand{\wt}{\widetilde}
\newcommand{\cA}{\mathcal{A}}
\newcommand{\g}{\gamma}
\newcommand{\G}{\Gamma}
\newcommand{\al}{\alpha}
\newcommand{\be}{\beta}
\newcommand{\si}{\sigma}
\newcommand{\ep}{\epsilon}

\newcommand{\mf}{\mathfrak}
\newcommand{\mb}{\mathbb}
\newcommand{\p}{\partial}
\newcommand{\Del}{\nabla}
\newcommand{\nn}{\nonumber}
\newcommand{\vep}{\varepsilon}
\newcommand{\dl}{\bm{\delta}}
\newcommand{\cG}{\mathcal{G}}

\begin{document}
{\renewcommand{\thefootnote}{\fnsymbol{footnote}}

\begin{center}
{\LARGE  Gauge Gravity: a forward-looking introduction}\\
\vspace{1.5em}
Andrew Randono\footnote{e-mail address: {\tt arandono@perimeterinstitute.ca}}
\\
\vspace{0.5em}
The Perimeter Institute for Theoretical Physics \\
31 Caroline Street North\\
Waterloo, ON N2L 2Y5, Canada
\vspace{1.5em}
\end{center}
}

\setcounter{footnote}{0}

%%%%%%%%%%%%%%%%%%%%%%%%%%%%%
%%%%%%%%%%%%%%%%%%%%%%%%%%%%%

\begin{abstract}
This article is a review of modern approaches to gravity that treat the gravitational interaction as a type of gauge theory. The purpose of the article is twofold. First, it is written in a colloquial style and is intended to be a pedagogical introduction to the gauge approach to gravity. I begin with a review of the Einstein-Cartan formulation of gravity, move on to the Macdowell-Mansouri approach, then show how gravity can be viewed as the symmetry broken phase of an (A)dS-gauge theory. This covers roughly the first half of the article. Armed with these tools, the remainder of the article is geared toward new insights and new lines of research that can be gained by viewing gravity from this perspective. Drawing from familiar concepts from the symmetry broken gauge theories of the standard model, we show how the topological structure of the gauge group allows for an infinite class of new solutions to the Einstein-Cartan field equations that can be thought of as degenerate ground states of the theory. We argue that quantum mechanical tunneling allows for transitions between the degenerate vacua. Generalizing the tunneling process from a topological phase of the gauge theory to an arbitrary geometry leads to a modern reformulation of the Hartle-Hawking ``no boundary" proposal. 
\end{abstract}
 \pagebreak
 
\tableofcontents

\pagebreak
%%%%%%%%%%%%%%%%%%%%%%%%%%%%%
%%%%%%%%%%%%%%%%%%%%%%%%%%%%%

\section{Overview of the review}

This article is intended to be a pedagogical introduction to the gauge approach to gravity, with an emphasis on new insights and new lines a research that can be discovered by viewing gravity from this perspective. Briefly, the gauge approach to gravity is a reformulation, and in some respects a generalization, of the Einstein-Hilbert approach to general relativity that makes the gravitational interaction look more like the interactions that are familiar from the Standard Model of particle physics. The biggest hurdle to getting there is to recast the theory not as a theory of a dynamical {\it metric} as in Einstein-Hilbert gravity, but as a theory of a dynamical {\it connection} as in the Standard Model gauge theories. Once this is done, we can use some of the familiar concepts and tools of Standard Model physics and try to apply them to gravitational physics or cosmology. This is the ultimate goal of the article: to give a pedagogical introduction to this construction an see how it can be applied to get new things.

The review is geared toward graduate students or postdocs and faculty interested in modern formulations of gravity. As I mentioned, this review is intended to be primarily a pedagogical tutorial and an introduction to some new lines of research -- it is not a historical review. For a good review of the history of Einstein-Cartan gravity leading up to the Poincar\'{e} gauge theory see \cite{Hehl:1994ue}. I will  assume that readers are familiar with general relativity, and are comfortable enough with gauge theories to know, for example, that a connection, $A$, can be viewed (locally) as Lie algebra valued one-form. I will also assume some basic knowledge of differential geometry and group theory, on a practical level. This means understanding basic concepts such as differential forms and how to manipulate them, integration on manifolds, very basic topology, and Lie algebras and there relation to Lie groups. But, the focus will be on mathematical concepts and tools, not rigor. Recommended resources for the mathematical background are \cite{Baez:Book}\cite{Schutz:Book}\cite{Frankel}\cite{Nair:QFT}\cite{Eguchi:1980jx}.

Being rather long, let me begin with an overview of this review. Since most introductory texts \cite{Carroll}\cite{MTW}\cite{Wald} only focus on the second order Einstein-Hilbert approach to gravity, I begin with a review of Einstein-Cartan gravity in section \ref{ECG}. However, since this review will be brief, one may want to consult other sources (e.g. appendix of \cite{Carroll}, \cite{Baez:Book}\cite{Ortin}). This will culminate in a construction of the Einstein-Cartan action in section \ref{ECAction} and a discussion of how to couple spinors to gravity in the Einstein-Cartan framework, which will also serve to introduce the Clifford algebra representation I will use on and off for the rest of the article. Section \ref{GaugeGravity} synthesizes these ideas to take gauge approach to gravity one step further, by combining all the ingredients of Einstein-Cartan gravity into a single object. First I discuss in section \ref{CartanAlgebras} some of the mathematical foundations of this procedure, which falls into the framework of homogenous Klein geometries and reductive Cartan algebras. I then apply these ideas in section \ref{dSGroup} to show that de Sitter, anti de Sitter, and Minkowski space can all be thought of as homogeous Klein geometries. Following up, I show how the geometric content of these spacetimes can be encoded in a single, flat Cartan connection based on the (A)dS or de Sitter Lie group. Taking these ideas yet one step further, in section \ref{MMConstruction} I review the Macdowell-Mansouri action \cite{MMoriginal}, pointing out some of its deficiencies, namely its failure to be invariant under the full gauge group. Then drawing on analogy with the symmetry breaking mechanisms of the standard model, I review the Stelle-West model \cite{West:1978Lagrangian}\cite{Stelle:1979aj}\cite{Stelle:1979va} for a fully gauge invariant action with spontaneously broken symmetry. I conclude section \ref{GaugeGravity} with a discussion of some of the very peculiar features of the gauge formulation of gravity.

For the remainder of the paper I focus on the de Sitter case specifically. In sections \ref{WhydS} and \ref{GroupTopology} I show why of the three groups in consideration, the de Sitter group is so special. A full understanding of the implications of this requires an understanding of certain topological aspects of the gauge group, so I start with the simpler case of certain topological aspects of $SU(2)$ gauge theory in section \ref{WindingNumbers}. I then use these tools in section \ref{3Sphere} to construct an infinite class of exotic geometries on the three-sphere that will be relevant to the de Sitter gauge theory, and I explore these geometries in further detail in section \ref{NewGeometry}. 

After the brief detour into the geometries of the three sphere, I return to the de Sitter case in section \ref{dSExtension} where I apply the ideas of the previous sections construct an infinite class of de Sitter-like geometries. In section \ref{Interpretation}, I discuss the geometries in detail, focusing on the topological properties that distinguish each geometry from the next.

In the remainder of the article, section \ref{QuantumGravity}, I discuss aspects of the quantum theory of the gauge approach to gravity. Drawing from analogy with QCD and Yang Mills theories, I argue for the existence of an infinite class of quasi-stable ground states, semi-classically represented by the infinite class of de Sitter-like geometries. I then argue that it should be expected that quantum mechanical tunneling between the ground states allows for transitions between two inequivalent vacua known as instantons. Generalizing the tunneling process to arbitrary geometries can yield a modern reformulation of the Hartle-Hawking ``no boundary" proposal \cite{HawkingHartle}. Then, again following the analogy with QCD, in section \ref{Theta} I argue that the true stable quantum vacuum is not one of the concrete geometries discussed in the previous sections, but a coherent superposition of all such geometries known as a theta-state.

%%%%%%%%%%%%%%%%%%%%%%%%%%%%%
%%%%%%%%%%%%%%%%%%%%%%%%%%%%%

\subsection{Einstein-Cartan gravity \label{ECG}}
The quickest route to understanding gravity as a gauge theory is via Einstein-Cartan theory. Having a long history (see e.g. \cite{Hehl:1994ue}), this formulation of gravity goes by various combinations of the names Einstein, Cartan, Utimaya, Sciama, and Kibble \cite{Kibble:1961ba}\cite{Utiyama:1973nq,Utiyama:1980bp}. I will stick with the Einstein-Cartan moniker. Without embellishment, Einstein-Cartan theory is a slight tweak of Einstein-Hilbert gravity that is completely consistent with all the experimental tests of gravity. The main difference between the former and the latter is the allowance of torsion, which is then dynamically constrained by the matter content (see \cite{Hammond--Torsion} for a thorough review of torsion in gravitational theories). In the presence of fermions, the torsion is generically non-zero (albeit small) however it does not propagate on its own through the vacuum as it always vanishes when the fermionic fields vanish. From an observational perspective, in the testable realms of gravity achievable by contemporary experimental techniques, the presence of torsion in this form is undetectable. However, from a theoretical perspective it makes all the difference. Most modern formulations of quantum gravity (including String Theory, supergravity, and Loop Quantum Gravity) use the Einstein-Cartan formulation of gravity or its various offspring. The main advantage of the approach is that it allows for a closer parallel between gravity and the ordinary gauge interactions of the Standard model of particle theory.

I will now very briefly review some of the basic concepts. This is not intended to be an exhaustive review. In part, it will serve to establish some of the index and various other conventions I will use through the rest of these notes. 

The Einstein-Cartan formulation of gravity begin with a shift of focus from gravity as a dynamical theory of a metric, to gravity as a dynamical theory of something else, which I will now describe. The metric tensor, is a non-degenrate tensor, $\bm{g}$, that takes two vectors and spits out a real number: $\bm{g}(\bar{U},\bar{V}) \in \mb{R}$. In a coordinate  basis, the metric can be written in its usual component form
\beq
\bm{g}=g_{\mu\nu} \,dx^\mu \otimes dx^\nu\,.
\eeq
The equivalence principle says that we can always find a new set of coordinates $y^\mu$ such that at a chosen point, the metric looks like that of Minkowski space (and it will look approximately like Minkowski space in a small enough region surrounding this point). Thus in this frame, at this point (call it $\mathcal{P}$), the metric becomes
\beq
\bm{g}\mid_\mathcal{P} = g_{\mu\nu} \,dx^\mu \otimes dx^\nu \longrightarrow \bm{g}\mid_\mathcal{P} = \eta_{\mu\nu} \,dy^\mu \otimes dy^\nu
\eeq
where $\eta_{\mu\nu}=diag(-1,1,1,1)$. The presence of curvature means that we cannot extend this beyond the small neighborhood (technically even beyond the single point $\mathcal{P}$ except in approximation). However, nothing stops of from changing the set of basis one-forms $dy^\mu$, which are a coordinate basis, to a non-coordinate basis. In doing so, we can in fact ``trivialize" the metric components everywhere. Thus, let $e^{(\mu)}$ represents a set of four basis one-forms where for now the index ``$(\mu)$" just labels which of the four basis one-forms we are talking about. This basis is chosen so that {\it globally} we have\footnote{When I say globally this should be taken with a grain of salt. As is well known there are sometimes topological obstructions to finding a globally defined set of $n$-linearly independent vectors (or one-forms in our case). This can't be done on the two-sphere for example (you can't come the hair on a sphere (without leaving a bare spot somewhere)). Thus when I say globally it means as global as one can get subject to these topological obstructions.}
\beq
\bm{g}=\eta_{\mu\nu}\,e^{(\mu)} \otimes  e^{(\nu)}\,.
\eeq
Since this basis $e^{(\mu)}$ is special, it deserves a new set of indices. So I will introduces the indices $I,J,K,L,...$ taking values in $\{0,1,2,3\}$, and drop the parentheses. The new set of basis one-forms are now $e^I$. This set of fields goes by various names including tetrad, veirbein (meaning four legs), and co-frame (or sometimes just frame). The (true) frame is the set of dual vectors distinguished with a bar, $\bar{e}_J$ satisfying $e^I(\bar{e}_J)=\delta^I_J$, analogous to the vectors $\frac{\p}{\p x^\mu}$ dual to the coordinate basis $dx^\mu$. One easy way to distinguish coordinate from non-coordinate bases is to take the Lie bracket: $[\bar{e}_I , \bar{e}_J]\equiv \mathcal{L}_{{\bar{e}}{}_{I}} \bar{e}_J =- \mathcal{L}_{{\bar{e}}{}_{J}} \bar{e}_I$. If the basis is a coordinate basis, the bracket will always be zero. 

The key conceptual leap from Einstein-Hilbert gravity to Einstein-Cartan gravity, is to encode the spacetime dynamics not in the form of the components $g_{\mu\nu}$, but in the co-frame $e^I$. Of course this is required since in this orthonormal basis, the metric components $\eta_{IJ}$ are trivial. Thus, for example, when one writes down the Einstein Hilbert action, it is explicitly a functional of $g_{\mu\nu}$ (and its first derivatives, and second derivatives), whereas the Einstein-Cartan action is explicitly a functional of $e^I$ (and its first derivatives, but not its second derivatives, the reason for which will become more clear later). 

A few more words on the frame before we move on. The co-frame can be viewed as a map which take you from the tangent space $T_{\mathcal{P}}M$ to a new vector space $\mathbb{V}$ where the inner product on that vector space is more trivial. To see this take a vector field $\bar{V}$ and contract it with the co-frame $e^I(\bar{V})\equiv V^I$. The components $V^I$ are the components of the vector field in a basis where the norm is given by $\eta_{IJ}V^I V^J$. As always, one can always expand the co-frame in a coordinate basis as follows: $e^I= e^I_\mu \,dx^\mu$. The co-efficients themselves can be thought of as the map that takes vector components in the coordinate basis to the components in the orthormal basis since 
\beq
V^I=e^I(\bar{V}) =e^I ( V^\mu \frac{\p}{\p x^\mu})=e^I_\nu V^\mu \,dx^\nu(\frac{\p}{\p x^\mu})=e^I_\mu V^\mu\,.
\eeq
Thus $e^I_\mu$ plays the role of converting indices from coordinate to orthonormal bases. 

An observant reader may have noticed something peculiar here. Einstein-Cartan gravity is a dynamical theory of the coframe written in a coordinate basis as $e^I_\mu$, which has 16 independent components. On the other hand, Einstein-Hilbert gravity is a dynamical theory of the metric components $g_{\mu\nu}$, which only have 10 components. Where do the extra degrees of freedom come from and how could the two theories possibly be equivalent? This is where the magic of Einstein-Cartan gravity comes in. The extra six degrees of freedom are gauge degrees of freedom from a new gauge symmetry of Einstein-Cartan theory that is not present in Einstein-Hilbert gravity. The gauge group responsible for this new symmetry is the Lorentz group $SO(3,1)$, and this is the first step in recognizing gravity as a gauge theory. This will become more clear shortly. First, we will introduce the spin connection, which is the analogue of the connections describing gauge bosons of the standard model.

The remaining ingredient in Einstein-Cartan theory left to discuss is the connection defining parallel transport. To simplify the discussion, we begin with the standard Levi-Civita connection $\Gamma$ in a coordinate basis (our index conventions are such that $D^{(\Gamma)}_{\al}V^{\mu}=\p_\al V^\mu+\G^{\mu}{}_{\nu\al}V^\nu$)
\beq
\G^{\mu}{}_{\nu\al}=\G[g]^{\mu}{}_{\nu\al}=\frac{1}{2}g^{\mu\rho}\left(g_{\rho\nu,\al}+g_{\al\rho,\nu}-g_{\nu\al,\rho}\right)\,. \label{Levi-Civita1}
\eeq
Let's first recall how we got this formula. The Levi-Civita connection, expressed above as derivatives of the metric, is fixed uniquely (given the metric) by imposing two conditions. First, the connection is assumed to be symmetric, or torsion free. That is, the torsion {\it tensor} $T^\mu{}_{\nu\alpha}\sim\G^\mu{}_{\nu\alpha}- \G^\mu{}_{\al\nu}=0$. Second, the connection is assumed to be compatible with the metric in the sense that $D^{(\G)}_\alpha g_{\mu\nu}=0$. This allows us to solve for the connection to obtain (\ref{Levi-Civita1}). 

The next step enroute to Einstein-Cartan theory is to express the Levi-Civita connection in a more convenient way. Recall that the last index of the connection, the $\alpha$ in $\G^{\mu}{}_{\nu\al}$, transforms differently from the other indices in the sense that under a general coordinate transformation it transforms just like the component of a tensor (one-form in this case). The other two indices pick up an inhomogenous piece under the transformation. Thus, we will simply supress the indexes by contracting it with a basis one-form. The result can be thought of as a matrix valued one-form
\beq
\G^{\mu}{}_{\nu}\equiv \G^{\mu}{}_{\nu \alpha} \,dx^\al \,.
\eeq
Now, compare this connection to, for example, an $SU(2)$ connection of Yang-Mills theory (see e.g. \cite{Zee:QFT}\cite{Nair:QFT}) given by $A^{A}{}_B=A^i \frac{i}{2} \si^{iA}{}_{B}$ where $\si^{iA}{}_{B}$ are the ordinary Pauli matrices with matrix indices written explicitly. The connection takes values in the Lie algebra of the group, which for the Yang-Mills connection is just $\mf{su}(2)$. The same is true for the Levi-Civita connection, except that rather than $SU(2)$, the relevant group is the general linear group $GL(4,\mb{R})$. An element of the Lie algebra $\mf{gl}(4,\mb{R})$ is just an arbitrary $4\times 4$ real matrix denoted here by the $16$ index components of $\G^{\mu}{}_{\nu}$. The difference between the Levi-Civita connection and an arbitrary $SU(2)$ connection (aside from the gauge group) is that the Levi-Civita connection is highly constrained by the two conditions of metric compatibility and vanishing torsion. We can gain considerable insight into the nature of gravity as a gauge theory by casting these constraints in a different form, or getting rid of them altogether. 

Let's start with the metric compatibility condition. In fact this condition can be recast into a statement about the restriction of the gauge group to a natural subgroup of $GL(4,\mb{R})$. The trick is to write it in the orthonormal basis $e^I$. To do this, first recall that the coefficients $e^I{}_\mu$ are the maps from a coordinate frame to an orthonormal frame. But, they can also be thought of as almost arbitrary $4\times 4$ matrices, subject only to the condition that the determinant is not zero (one can show that $det(e)=\pm \sqrt {det(|g|)} \neq 0$). But this is precisely the condition defining an element of $GL(4,\mb{R})$. Now recall how the connection transforms under a arbitrary element $g$ (not the metric) of the gauge group. Supressing all indices we have
\beqa
A &\longrightarrow &  ^gA=gAg^{-1}-dg g^{-1} \quad \text{for} \quad g\in SU(2)  \nn\\
&\text{and}&  \nn\\
\G &\longrightarrow & ^g\G=g\G g^{-1}-dg g^{-1} \quad \text{for} \quad g\in GL(4,\mb{R})
\eeqa
Now, take $g=e \in GL(4,\mb{R})$. The inverse $e^{-1}$ in components is denoted $e^\mu{}_I$. This transforms the connection to the orthonormal basis. With indices this looks like
\beq
\G^{\mu}{}_{\nu} \longrightarrow \G^I{}_J =e^I{}_\mu \G^{\mu}{}_\nu e^\nu{}_J- d e^I{}_\rho\, e^{\rho}{}_{J} \,.  \label{ChristTransformation}
\eeq
The new connection (not really new, just written in a different basis) defines the covariant derivative of vectors living in the orthonormal vector space $\mb{V}$, by $D^{(\G)}V^I=dV^I +\G^I{}_J V^J$. The relation above is often written in another (for example in \cite{Carroll}), potentially confusing, way (which nevertheless can be useful, if only as a mnemonic device). Suppose we had a connection with associated covariant derivative $\Del$ that acted on both types of indices separately. Then the covariant derivative of the coframe components is
\beq
\Del_{\mu}e^I{}_\nu=\partial_\mu e^I{}_\nu +\Gamma^I{}_{K\mu} \,e^K{}_\nu-\Gamma^{\alpha}{}_{\nu\mu} \,e^I{}_\alpha=0 \,,
\eeq
which can be rearranged (inverting a tetrad here and there) to give (\ref{ChristTransformation}).

Now, consider the metric compatability condition in this basis (which still holds, since all we have done is transformed to a new basis):
\beqa
D_\G {g_{IJ}}&=& D_\G {\eta_{IJ}} \nn\\
&=& d\eta_{IJ} -\G^K{}_I \eta_{KJ} -\G^K{}_J \eta_{IK} \nn\\
&=& -\G_{JI} -\G_{IJ} \nn\\
&=& 0\,.
\eeqa
Thus, we have $\G^{IJ}=-\G^{JI}$, or $\G^{IJ}=\G^{[IJ]}$. This condition effectively reduces number of index components from $16$ to $6$. Moreover, these indices should indicate that the connection lives in the Lie algebra of some group. A $4\times 4$ matrix $\lambda^I{}_J$ that satisfies $\lambda^{IJ}=\lambda^{[IJ]}$ (where indices are raised using $\eta^{IJ}$) is an element of the Lie algebra $\mf{so}(3,1)$. Thus, the metric compatibility condition has reduced the gauge group from $GL(4,\mb{R})$ to $SO(3,1)$! 

In retrospect this is not terribly surprising...after all, we are simply restricting the set of general linear transformations that we can make to the the set of general linear trasnformations that preserve the form $\eta_{IJ}$. But this is precisely the subgroup $SO(3,1)\subset GL(4,\mb{R})$. One further comment is in order. Technically we have not uniquely fixed the gauge group, but simply the Lie algebra, since there can be more than one group associated with the Lie algebra. In fact, in the future we will take the gauge group to be the double cover of $SO(3,1)$, namely $Spin(3,1)\simeq SL(2,\mb{C})$.

We can now address the second condition that makes the Levi-Civita connection different from an arbitrary gauge connection, the condition of vanishing torsion. This condition we will simply relax. To do this recall that, being an affine space, all connections in the space of connections can be connected by adding a tensor (with indices in the right spots, the key being that the object transforms homogenously under a gauge transformation unlike the full connection). Thus given an {\it arbitrary} $SO(3,1)$ connection $\omega^I{}_J$, we can always express 
\beq
\omega^{I}{}_J=\G^I{}_J +C^I{}_J \,.
\eeq 
The tensor $C^I{}_J$ is known as the contorsion tensor. To relate the arbitrary connection to the Levi-Civita connection in the presence of torsion, we first note the identity 
\beqa
T^I &=&\frac{1}{2} T^I{}_{\mu\nu} dx^\mu \w dx^\nu \sim e^I{}_\al\left(\omega^\alpha{}_{\mu\nu}-\omega^\alpha{}_{\nu\mu} \right) \nn\\
&=& D_\omega e^I =de^I +\omega^I{}_J \w e^J\,.
\eeqa
The contorsion tensor therefore satisfies $T^I=C^I{}_J \w e^J$ since $D_{\G}e^I=0$. And it can be solved entirely in terms of the torsion to give
\beq
C_{IJK}=\frac{1}{2}\left(T_{KIJ}-T_{JKI}-T_{IJK} \right)\,.
\eeq

For future reference, the curvature of the Levi-Civita connection in the orthonormal basis, is clearly just the gauge transform of the ordinary curvature tensor to an orthonormal basis, since the curvature transforms like an ordinary tensor. The only possibly new thing is the way we will write the curvature. Since $\G^{IJ}$ and $\omega^{IJ}$ are one-forms valued in the Lie algebra of the gauge group ($\mf{spin}(3,1)$) the curvature is a two-form valued in the Lie algebra (the odd mix of coordinate and orthonormal indices may seem strange at first sight, but it is natural from the perspective of the fiber bundle construction):
\beqa
R_\G{}^{I}{}_{J} &=& d\G^{I}{}_J+\G^I{}_K \w \G^K{}_J \nn\\
&\equiv& R{}^{I}{}_{J\al \be} \,\frac{1}{2} dx^\al \w dx^\be  \nn\\ 
&=& e^I{}_{\mu} e^{\nu}{}_J\, R^{\mu}{}_{\nu\al\be} \,\frac{1}{2} dx^\al \w dx^\be\,.
\eeqa
The curvature of the spin connection can then be related to the curvature of the Levi-Civita connection and the contorsion by
\beq
R_\omega{}^{I}{}_J=R_\G{}^{I}{}_J + D_{\G} C^I{}_J + C^I{}_K \w C^K{}_J \,.
\eeq

We now have the two ingredients of Einstein-Cartan gravity. These are the coframe $e^I$ and what we will refer to as a spin-connection $\omega^I{}_J$. These will be the new dynamical ingredients describing gravity.

Let's now pause to see what we've done. In fact, quite alot. In retrospect, the presentation could have began like this (see \cite{Ashtekar:LQGReview1}\cite{Baez:Book}\cite{Carlip:Book})...consider a principle $G$-bundle with base manifold $M$, where $G$ is the gauge group $SO(3,1)$, and an associated vector bundle with typical fiber being the $SO(3,1)$ representation space $\mb{V}$. The Cartan-Killing form on $SO(3,1)$ induces an inner product on the vector bundle given by $\langle U | V\rangle =\eta_{IJ}U^I V^J$. Define a connection on the $G$-bundle identified in a local trivialization with the connection coefficients $\omega^I{}_J$. The connection is naturally compatible with the Cartan-Killing form in the sense that $D_\omega \eta_{IJ}=0$. Now consider an invertible and differentiable map $e:TM\rightarrow \mb{V}$. In a local trivialization of the vector bundle this map is given by a one-form $e^I$ taking values in the associated $SO(3,1)$ vector space $\mb{V}$. Taking the inner product of the map we can define $\bm{g}=\langle e |e \rangle =\eta_{IJ}\,e^I \otimes e^J$, which induces a metric in $T_*M\otimes T_*M$.

Either way we approach the construction, the lesson to be taken from it is that the dynamics of gravity has been recast into the form of a gauge interaction where the new ingredients describing the interaction are the tetrad $e^I$ and the spin-connection $\omega^I{}_J$.

\subsection{The Einstein-Cartan action \label{ECAction}}
Let us now turn to the action describing the gravitational interaction. As before, we can build this from the known form of the Einstein-Hilbert action (see \cite{MTW}). In the presence of a cosmological constant, $\Lambda$, the action is (with $k=8\pi G$)
\beq
S_{EH}=\frac{1}{2k} \int_M (Ricci-2\Lambda) \sqrt{|g|} d^4x \,.
\eeq
I will now rewrite this action in a less familiar form to ease the transition to the Einstein-Cartan action. First, define the densitized Levi-Civita alternating symbol by $\vep_{\mu\nu\al\be}=\sqrt{|g|} \ep_{\mu\nu\al\be}$, where $\ep_{\mu\nu\al\be}$ is just the ordinary completely anti-symmetric, alternating symbol with $\ep_{0123}=1$. The metric volume form is then given by $\wt{\si}=\sqrt{|g|} dx^0 \w dx^1\w dx^2 \w dx^3= \frac{1}{4!} \vep_{\mu \nu \alpha \beta}\,dx^\mu\w dx^\nu \w dx^\al \w dx^\be$, which replaces the $\sqrt{|g|} d^4x$ in the action. Now let's express the action as a function of the Levi-Civita curvature viewed as a $\mf{gl}(4,\mb{R})$ valued two-form. The reader can check that the action is equivalent to 
\beq
S_{EH}=\frac{1}{4k} \int_M \left( \vep_{\mu\nu\al\be} \,dx^\mu \w dx^\nu \w R_\G{}^{\al \be} -\frac{\Lambda}{6} \vep_{\mu\nu\al\be} \, dx^\mu \w dx^\nu \w dx^\al \w dx^\be  \right)\,.
\eeq
It may seem like overkill to express the action in this form, but the transition to the Einstein-Cartan action is now trivial. Since all the terms are now $GL(4,\mb{R})$ invariant, we can express the action in any basis, not just a coordinate basis. Thus, I choose to the orthonormal basis, so $dx^\mu \rightarrow e^I$, and the action becomes (grouping the two terms together and recognizing that in an orthonormal basis $\vep_{IJKL}= \ep_{IJKL}$ since $\sqrt{|g|}=\sqrt{|\eta|}=1$)
\beq
S_{EH}=\frac{1}{4k} \int_M \ep_{IJKL} \,e^I \w e^J  \w \left( R_\G{}^{KL} -\frac{\Lambda}{6} \,e^K \w e^L  \right)\,.
\eeq
This is still just the Einstein-Hilbert action, just written in an unfamiliar form in an orthonormal basis. The next step couldn't be simpler: just replace the Levi-Civita curvature $R_\G{}^{IJ}$ with $R_\omega {}^{IJ}$. This is the Einstein-Cartan action:
\beq
S_{EC}=\frac{1}{4k} \int_M \ep_{IJKL} \,e^I \w e^J  \w \left( R_\omega{}^{KL} -\frac{\Lambda}{6} \,e^K \w e^L  \right)\,.
\eeq

This last step may have seemed trivial, but there is more to it than first appears. First, the Einstein-Hilbert action is taken to be a functional of the metric alone, or the tetrad alone in the orthonormal basis, since the Levi-Civita connection in $R_\G$ can be expressed in terms of the metric/tetrad. On the other hand, in the Einstein-Cartan action, the tetrad $e^I$ and the spin connection $\omega^{IJ}$ can be taken to be genuine independent variables (the formula $\omega^{I}{}_J=\G[e]^{I}{}_J+C^{I}{}_J$ just shifts the independence to the contorsion, but one could equally well just forget this formula and think of $\omega^{I}{}_J$ itself as completely independent of $e^I$). One consequence of this is that whereas the Einstein-Hilbert action is second order in derivatives of the dynamic variables (metric or tetrad), the Einstein-Cartan action is first order in derivatives of $e$ and $\omega$. For this reason the Einstein-Cartan action is often referred to simply as the first order formulation of gravity.

To see the real difference between the two actions, let's look at the equations of motion. These are obtained by taking arbitrary variations $\dl e^I $ and $\dl \omega^{J}{}_K$ and setting the variation equal to zero (here and throughout I will ignore boundary terms):
\beqa
\dl S_{EC}&=& \frac{1}{4k}\int_M  \epsilon_{IJKL} \,\dl e^I \w \left(2\,e^J \w \left( R_\omega{}^{KL}-\frac{\Lambda}{3} \,e^K \w e^L \right)\right)  \nn\\
& & \quad + \frac{1}{4k} \int_M \epsilon_{IJKL}\,\dl \omega^{IJ} \w \left( D_{\omega} (e^K \w e^L) \right) \,.
\eeqa
With the addition of a matter action, setting the variation of the total action to zero, the equations of motion that emerge are
\beqa
\epsilon_{IJKL}\,e^J \w \left( R_\omega{}^{KL}-\frac{\Lambda}{3} e^K \w e^L \right)&=& -2k \,\frac{\dl S_{matter}}{\dl e^I}  \label{EC1}\\
\epsilon_{IJKL}\,D_{\omega} (e^K \w e^L) &=& -4k \,\frac{\dl S_{matter}}{\dl \omega^{IJ}}\,.\label{EC2}
\eeqa
When the spin-current density $\frac{\dl S_{matter}}{\dl \omega^{IJ}}$ is zero, and the tetrad is assumed to be invertible (which it almost always is, though we will relax this condition later), equation (\ref{EC2}) can be inverted to give $T^I=D_\omega e^I=0$. Thus, the vanishing torsion condition is achieved dynamically (in some cases) in Einstein-Cartan gravity. In these cases, the remaining equation can be shown to be exactly equivalent to the ordinary Einstein equations $G_{\mu \nu}=8 \pi G \,T_{\mu\nu}$. However, even when the spin-current is not zero (as it is when you try to couple fermions to gravity), the torsion equation (\ref{EC2}) is an {\it algebraic} equation (as opposed to a differential equation). This means that the torsion is completely determined by the matter content, and does not have dynamical degrees of freedom that exist on its own. For this reason, people say that torsion is non-propagating in Einstein-Cartan theory (it cannot propagate through empty space on its own like, say, gravitational waves can). 

\subsection{Coupling to spinors and the Clifford algebra notation \label{Spinors}}
One of the main advantages of Einstein-Cartan gravity is that it allows a simple coupling of gravity to spinors. At a fundamental level, a Dirac spinor $\psi$ here viewed as a complex four-component object, is an object living in the fundamental representation of the double cover of the Lorentz group, $\overline{SO}(3,1)=Spin(3,1)$. In terms of $Spin(3,1)$, the tetrad $e^I$ is a one-form taking values in the adjoint (vector) representation, and $\omega^{I}{}_J$ is a $Spin(3,1)$ connection in the adjoint representation. To couple to spinors, we need to transform these variables to the fundamental representation. To do this it is useful to introduce the Clifford algebra representation of $Spin(3,1)$. This will also serve as a segue into the next sections where we will use this notation extensively. 

The Clifford algebra is the algebra of (usually matrices) $\g^I$ defined by the condition
\beq
\g^I \g^J +\g^J \g^I =2\eta^{IJ}\, \bf{1}\,.
\eeq
For our purposes, the gamma matrices can be used to build various Lie algebras. Most importantly for now, the Lie algebra $\mf{spin}(3,1)\simeq \mf{so}(3,1)$ is spanned by the six ``bivector" elements\footnote{Note: when contracting anti-symmetric objects, we usually add an extra factor of $\frac{1}{2}$ to avoid overcounting. Thus for example, I will use $\lambda=\frac{1}{4}\g^{[I}\g^{J]} \,\lambda_{IJ}$ to convert between the fundamental and adjoint representations.} $\frac{1}{2} \g^{[I}\g^{J]}$. These naturally act on the spinor $\psi$ and exponentiating them gives the fundamental representation of $Spin(3,1)$. In the fundamental representation, the spin-connection is a one-form that takes values in the bivector elements of the Clifford algebra:
\beq
\omega\equiv \omega_{IJ}\,\ts{\frac{1}{4}}\g^{[I}\g^{J]} \,.
\eeq
When working in the fundamental representation I will generally drop all indices as I have done above. Thus for example, the exterior covariant derivative of a spinor is $D_\omega \psi =d\psi +\omega \psi $. The curvature is also Lie algebra valued, and it looks like this in our index free notation:
\beq
R_\omega=d\omega+ \omega \w \omega =(d\omega^{IJ} +\omega^{I}{}_K \w \omega^{KJ}) \ts{\frac{1}{4}} \g_{[I} \g_{J]} =R_{\omega}{}^{IJ}\, \ts{\frac{1}{4}} \g_{[I} \g_{J]}\,.
\eeq

The tetrad $e^I$ is then naturally valued in the vector elements $\frac{1}{2}\g^I$ (with normalization chosen for future convenience) so that, again using our index free notation, $e\equiv e^I \,\frac{1}{2}\g_I$. The exterior derivative of the tetrad is then given by
\beq
D_{\omega}e=de+\omega \w e +e\w \omega =(de^I+\omega^I{}_J \w e^J) \ts{\frac{1}{2}} \g_I 
\eeq
which, of course, can be identified with the torsion $T \equiv T^I \, \frac{1}{2} \g_I$. 

The trace properties of the Dirac matrices should be familiar from Quantum Field Theory, so this allows for an aesthetically pleasing form for the action:
\beq
S_{EC}=\frac{1}{k} \int_M Tr \left(\star \,e \w e\w \left(R_\omega -\frac{\Lambda}{6} \,e\w e \right)\right)
\eeq
where $\star\equiv -i\g_5 =\g^0 \g^1\g^2\g^3=\frac{1}{4!}\epsilon_{IJKL} \g^I\g^J\g^K\g^L $ acts in the fundamental representation like the dual operator $\epsilon_{IJKL}$ did in the adjoint representation. To make this action even easier on the eye, I will adopt the (probably non-standard) habit of dropping the explicit trace, and the explicit wedge products between differential forms, when it is obvious that they should be there. So, the action now looks like 
\beq
S_{EC}=\frac{1}{k}\int_M \star \,e\,e\,R_\omega -\frac{\Lambda}{6} \,\star e\,e\,e\,e\,.
\eeq

%%%%%%%%%%%%%%%%%%%%%%%%%%%%%
%%%%%%%%%%%%%%%%%%%%%%%%%%%%%

\section{Gauging Gravity \label{GaugeGravity}}
The Einstein-Cartan formulation of the previous sections provides the first step to realizing gravity as a gauge theory.  We have made some progress into placing the gravitational interaction on the same (or similar) footing as the interactions of the standard model. Both theories are based on a connection over a principle $G$-bundle, the gauge bosons (the gluons, W, and Z) of the standard model being analogous to the spin connection $\omega$. On the other hand, there are some obvious differences. The glaring difference is the existence of a new field $e$, which serves as a map from the tangent space to the $SO(3,1)$ representation space and imbues spacetime with its metric structure. The kinetic term of gravity (i.e. the Einstein-Cartan action terms not involving matter fields), which should be the analog of $\int *F\w F$ for the gauge bosons, looks completely different in form. 

This is where the next step begins. The basic idea is to incorporate the frame field $e$ and the spin connection $\omega$ into a single connection based on a larger gauge group. Since the six-dimensional Lorentz group should be a subgroup of the new gauge group, and the tetrad has $4$ internal degrees of freedom, we should expect the larger group to be at least $10$-dimensional. In fact there is a very natural mathematical way to do this. This unification falls into the category of reductive Cartan algebras, which I will briefly review now (for a more thorough and better introduction, see \cite{Wise:2009fu}\cite{Wise:MMgravity}).

\subsection{Reductive Cartan algebras and homogenous Klein geometries: Poincar\'{e} gauge theory \label{CartanAlgebras}}
Imagine we have a highly symmetric space $X$ that will act like a preferred ``ground state" of our gauge theory. The symmetry of the space means that there is a group $G$ that acts transitively on the manifold such that every point on the manifold can be obtained by from any other by the group action. Now suppose there is a subgroup $H$ that preserves some point, say, $x\in X$. Since all points of the manifold can be obtained by applying different group elements $g\in G$, but $h\in H\subset G$ preserves the point, the set of points of the homogenous space is in one-to-one correspondence with the coset space $G/H$. The coset space is referred to as a homogenous Klein geometry, and it will serve as our model space. 

To obtain the relevant homogenous Klein geometries appropriate for a Lorentzian signature metric in four dimensions, we will work backwards. Recall that in four dimensions, a maximally symmetric geometry has $10$ Killing vectors (there is that number again), corresponding to the $3$ rotations, $3$ boosts, and $4$ transvections (generalizations of translations to curved spaces). This restriction of maximal symmetry is highly constraining: in fact there are only three choices corresponding to zero, constant negative, and constant positive curvature. These geometries are the well known geometries corresponding to Minkowski space ($0$), anti-de Sitter space ($-$), and de Sitter space ($+$). 

Let's look at Minkowski space, since it is the simplest to visualize. The symmetries of Minkowski space consist of the set of rotations, boosts, and translations, and together they constitute the Poincar\'{e} group denoted\footnote{The semi-direct product $\ltimes$ can be understood easily at the level of the Lie algebra. If we denote $\mf{h}=\mf{so}(3,1)$ and $\mf{p}=\mb{R}^{3,1}$, the Lie algebra schematically satisfies $[\mf{h},\mf{h}] \subseteq \mf{h}$ and $[\mf{p}, \mf{p}]=0$ (translations commute). But $[\mf{h},\mf{p}]\subseteq \mf{p}$ since a rotation or boost of a translation is still a translation. Thus, the group is not a simple direct product of $SO(3,1)$ and $\mb{R}^{3,1}$, but the next simplest thing, namely the semi-direct product of the two.} $ISO(3,1)=SO(3,1)\ltimes \mb{R}^{3,1}$. In this case, the group is $G=ISO(3,1)$, and the stabilizer is the subgroup that preserves one point (which might as well be what we call the origin), $H=SO(3,1)$. In this case the coset space is $ISO(3,1)/SO(3,1)=\mb{R}^{3,1}$, which can be identified with Minkowski space.

Now, the rough idea is to supplement the tangent space at each point $T_x M$ with the homogenous space $X$. The two spaces have the same dimension, and one can imagine a map taking one to the other. In the Poincar\'{e} case, the tangent space is supplemented with the affine space $\mb{R}^{3,1}$, so that we are free to slide around the point of contact between the manifold and $X$. A Cartan-connection is then just an ordinary connection, represented in a local trivialization, by the coefficients $\cA$, which as usual is a one-form now valued in the Lie algebra $\mf{iso}(3,1)$. Corresponding to the identification of the subgroup $H$ in the Lie group $G$, we can decompose the Lie algebra into the subalgebra $\mf{h} \subset \mf{g}$ and its complement $\mf{p}=\mf{g}/\mf{h}$. The boost and rotation generators of $\mf{so}(3,1)$ are in $\mf{h}$ whereas the translations are in $\mf{p}$. Thus corresponding to $\mf{g}=\mf{h}\oplus \mf{p}$, the connection decomposes into 
\beq
\cA=\omega +\frac{1}{\ell}e\,.
\eeq
I have chosen the symbols suggestively here. The parameter $\ell$ is just an arbitrary parameter with dimensions of length. The fundamental idea of the Cartan approach to connections is to identify the spin connection itself with the $\mf{h}$-component of the connection, and the coframe with the $\mf{p}$-component. This is made possible by the fact that the subalgebra $\mf{p}$ has the same dimension as the manifold. Because of this, the one form $e$ can be thought of as a map from the tangent space $TM$ to the tangent space of the homogenous Klein geometry $G/H$ serving as the model space for the gauge theory.

The idea of extending the connection describing the gravitational interaction to a connection valued in $\mf{iso}(3,1)$ with the spin connection emerging as the Lorentz piece and the coframe as the translation piece goes by the name of Poincar\'{e} gauge theory in physics. This has been explored extensively (see \cite{Hehl:1994ue} for a review). In the presence of a  non-zero cosmological constant, the procedure can be extended to the de Sitter and anti-de Sitter groups. This will be the focus of this review, with a special interest in the de Sitter case for reasons that will become clear shortly.

\subsection{Extension to de Sitter and anti-de Sitter \label{dSGroup}}
The remaining two maximally symmetric geometries one can put on a four-manifold are described by de Sitter space and anti-de Sitter space. The quickest route to understanding these spaces and their isometries is to embed them in a larger $5$ dimensional space (see \cite{Hawking:LSS}\cite{Moschella:deSitter}). 

\begin{figure}
 \begin{center}
\includegraphics[height=5.0cm]{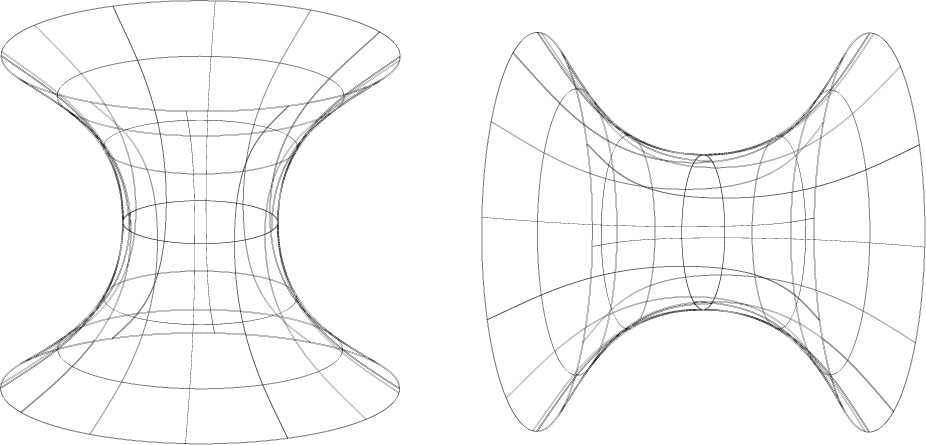}
   \end{center}
  \caption{\label{dSAdS} On the left is the canonical picture of de Sitter space viewed as a hyperboloid immersed in a five-dimensional Lorentzian signature space. Each horizontal slice of the hyperboloid represents a three-sphere that starts large, contracts to a minimum at the throat, and expands again in a time symmetric way. Anti-de Sitter space (right) can be pictured as a hyperboloid turned on its side, but immersed in a $(-,+,+,+,-)$ space. However, it should be understood that usually when one talks about AdS spacetime, they mean the spacetime formed by cutting and unwrapping the hyperboloid and gluing copies together to form the universal cover.
  }
  \end{figure}

On the five dimensional flat space take the metric to have Lorentz signature $(-,+,+,+,+)$. Now, consider the hyperboloid (see Fig. (\ref{dSAdS}) defined by the ``constant radius" condition (here indices $A,B,C,...$ to values $\{0,1,2,3,4\}$)
\beq
\eta_{AB} \,X^{A} X^{B} =-dT^2+dX^2+dY^2+dZ^2+dW^2=\ell^2\,.
\eeq

The pull-back of the metric $\eta_{AB}$ to the hyperboloid defines a constant curvature geometry on a manifold with topology $\mb{R}\times \mb{S}^3$. More specifically, given a tetrad $e^I$ such that $\mf{g}=\eta_{IJ} e^I\otimes e^J$, is the pull-back of the metric to the hyperboloid, the Levi-Civita connection $\G=\G[e]$ satisfies
\beq
R_\G^{IJ}=\frac{\Lambda}{3} \,e^I \w e^J  \quad \quad T^I=0 \label{CCcondition}
\eeq
where $\Lambda >0$ is the positive cosmological constant and is related to the length parameter by $\ell=\sqrt{\frac{3}{\Lambda}}$.

Similarly, anti-de Sitter space can be defined by choosing the metric to have signature $(-,+,+,+,-)$, and taking the the ``constant radius" space
\beq
\eta_{AB} X^A X^B=-dT^2 +dX^2+dY^2 +dZ^2-dW^2=-\ell^2
\eeq
This space again satisfies the constant curvature condition (\ref{CCcondition}) with $\Lambda <0$. However, defined as such, the space has closed timelike curves (picture the hyperbola turned on it side so that one of the compact dimensions lies along the $T$-axis). To overcome this problem, one can cut open the space, unfold it, and glue copies together along the cut. This effectively unwraps the $\mb{S}^1$ of the manifold with topology $\mb{S}^1\times \mb{R}^3$ to turn it into a manifold $\mb{R}^4$ with constant negative curvature. This procedure is known taking the universal cover of the manifold, and when people refer to anti-de Sitter space, they are usually referring to this universal covering space.

The advantage of defining the spaces like this is that the isometries of the spaces are transparent. The isometries are defined by the set of transformations we can make that preserve the metric and the defining conditions of the two spaces. These transformations are clearly $SO(4,1)$ for de Sitter, and $SO(3,2)$ for anti-de Sitter (i.e. they consist of the boosts and rotations that preserves the flat metric on the $5$D space, but not the translations since these will move the embedded constant radius subspaces). So we have
\beqa
G=\begin{cases} ISO(3,1) & \mbox{for Minkowski space, } M_4\  (\Lambda=0) \\
SO(4,1) &\mbox{for de Sitter space, } dS_4\ (\Lambda>0)  \\
SO(3,2) & \mbox{for anti-de Sitter space, } AdS_4\ (\Lambda<0) \,.
\end{cases}
\eeqa
In each of these cases, the stabilizer group (i.e. the group that holds fixed some point that we call the origin) is the Lorentz group $SO(3,1)$. In all three cases, the maximally symmetric geometry can be identified with the homogenous Klein geometry
\beq
\mbox{Homogenous Klein geometry } G/H=\begin{cases} ISO(3,1)/SO(3,1)  = & M_4 \ \ (\Lambda=0) \\
SO(4,1)/SO(3,1)  = & dS_4 \ \ (\Lambda >0) \\
SO(3,2)/SO(3,1)  = & AdS_4 \ \ (\Lambda<0)\,.
\end{cases}
\eeq

Now let's look at the Lie algebras. In all cases, the Lie algebra splits into a direct sum of the stabilizer subalgebra and the transvections, $\mf{g}=\mf{h}\oplus \mf{p}$, where $\mf{h}=\mf{so}(3,1)$. In general, the coset space structure implies that the algebra has the generic form
\beq
[\mf{h},\mf{h}]\subseteq \mf{h}  \quad \quad [\mf{h}, \mf{p}] \subseteq \mf{p}\,.
\eeq
This is true of all reductive Cartan algebras. In addition to this, each of the groups above has the additional property that they are {\it symmetric}, which means that there is an involution operations that allows one to grade the algebra such that $\mf{h}$ are the ``even" elements, and $\mf{p}$ are the ``odd" elements. This is not true of all reductive Cartan algebras, but it holds for all the cases we will concern ourselves with. It implies the more restrictive condition
\beq
[\mf{p},\mf{p}]\subseteq \mf{h}
\eeq
since the left hand side, involving the product of two odd elements, must be even. 

One more comment before we move on. In the discussion above we have focused on the isometry group of the homogenous spaces, for which it was sufficient to work with the orthogonal groups $SO(m,n)$. But, the essence of the discussion is independent of whether we use the orthogonal groups, or their double cover $\overline{SO}(m,n)=Spin(m,n)$. This distinction is essential when coupling gravity to spinors, but it will also play a major role in what follows. There are various identifications, one can make with these gauge groups, which may or may not be illuminating. The most important ones are $Spin(3,1)\simeq SL(2,\mb{C})$, $Spin(4,1)\simeq Sp(2,2,\mb{R})$, and $Spin(3,2)\simeq Sp(4,\mb{R})$. The last isomorphism plays a fundamental role in the construction of supergravity, as $Sp(4,\mb{R})$ can be viewed as the group preserving a spinorial inner product (which looks symplectic when restricted to Majorana spinors), and it is the reason why supergravity prefers a negative cosmological constant \cite{Ortin}.

Let's focus on the de Sitter or anti-de Sitter cases now. For the connection it is useful to fix the $SO(4,1)$ or $SO(3,2)$ gauge such that the the indices $I,J,K,...$ taking values in $\{0,1,2,3\}$ are indices in the $H=SO(3,1)$ representation space. Then, we can make the identification 
\beq
\cA^A{}_B \longrightarrow \begin{cases} A^I{}_J  =\omega^I{}_J \\ A^I{}_4=\frac{1}{\ell}e^I\,.\end{cases}
\eeq
The curvature of the connection also splits into components as (the $-$ is for de Sitter and the $+$ for anti-de Sittter)
\beq
F_\cA^{AB}\ \longrightarrow\  \begin{cases} F^{IJ}=d\omega^{IJ}+\omega^{I}{}_K\w \omega^{KJ}\mp \frac{1}{\ell^2} \,e^I\w e^J =R^{IJ}\mp \frac{1}{\ell^2} \,e^I \w e^J \\ F^{I}{}_4=\frac{1}{\ell}\left(de^I +\omega^I{}_K\w e^K\right) =\frac{1}{\ell}\,T^I \,.
\end{cases}
\eeq
The first line above is the $\mf{h}$-valued part of the curvature, sometimes referred to as the corrected curvature, and the second line is the $\mf{p}$ valued part of the curvature. This gives a new meaning to the torsion -- torsion is simply one component of the curvature of the Cartan connection. It is stable under $H$, meaning it transforms like a vector, but in general a gauge transformation in $G$ will mix the torsion and the corrected curvature. As you might guess, Similar properties hold for the Poincar\'{e} case, but there the constant $\ell$ does not have a clear physical meaning since it is not related to the cosmological constant.

One particularly convenient, and interesting, relation emerges from the identifications above. Suppose the tetrad is everywhere nondegenerate and we have fixed the topology of the manifold to be that of the homogenous Klein geometry. Then in every case, the homogenous Klein geometry, (i.e. $M_4$, $dS_4$, or $AdS_4$) is the unique solution (subject to those conditions) satisfying
\beq
F_\cA^{AB}=0\,.
\eeq
This is what we meant when we said that the homogenous Klein geometry will serve as a model for the ``ground state" of the gauge theory -- the Klein geometry is the nondegenerate, flat curvature solution (provided we have chosen the topology properly). This property will be very important in later sections.

Above we have given the form of the connection in the fundamental representation of $SO(m,n)$, or the adjoint representation of $Spin(m,n)$. Using the Clifford algebra, we can represent the connection in the fundamental representation of $Spin(m,n)$. The spin connection $\omega$ as we have seen is valued in the bivector elements of the Clifford algebra, which form a representation of the Lie algebra $\mf{h}=\mf{so}(3,1)$. In each of the three cases, the transvections can be represented by some linear combination of $\g^I$ and $\g_5 \g^I$. To help keep track of minus signs, I will stick with the convention that $e$ is always defined by $e=e^I \,\frac{1}{2}\g_I$, and the $\g_5$ factors I will pull out of the frame when writing the connection. The following identifications work:
\beq
\mf{h}=span\{ \frac{1}{2}\g^{[I}\g^{J]} \} \quad \quad \mf{p} =\begin{cases} span\{\frac{1}{2}(1+\g^5) \g^I \} &\mbox{ for } \Lambda=0\\
span\{ \frac{1}{2}\g_5 \g^I \} &\mbox{ for } \Lambda>0\\
span \{ \frac{1}{2} \g^I \} &\mbox{ for } \Lambda<0 \end{cases} 
\eeq
Thus the Cartan connection split into 
\beq
\cA = \begin{cases}  \omega + \frac{1}{2\ell}(1+\g_5) e  &\mbox{ for } \Lambda=0 \\
\omega+\frac{1}{\ell} \g_5 e &\mbox{ for } \Lambda>0 \\
\omega+\frac{1}{\ell} e &\mbox{ for } \Lambda<0 \end{cases}
\eeq
The normalization is chosen such that in all three cases, the homogenous ground state is given by the (nondegenerate) flat connection satisfying
\beq
F_\cA=0 \longrightarrow \begin{cases} R_\omega= \frac{\Lambda}{3} \,e \w e \\ T=0 \end{cases}
\eeq

\subsection{The Macdowell-Mansouri mechanism/construction \label{MMConstruction}}
Now let's return to the action. Recall the gravity action, $\int \star \,e\,e\,R$, looks very different from the kinetic term of the standard model gauge bosons, $\int * F \,F$. Is there a way we can make it better? As we have seen in the last section, the spin connection and the tetrad combine into a single connection $\cA$. Maybe the curvature of that connection plays a more fundamental role, and we should try building an action with it. The way to do this was first noticed by Macdowell and Mansouri in a seminal paper \cite{MMoriginal}.

The most natural thing to do might seem to write down the action $\int *F_\cA \,F_\cA$ but closer inspection shows that there are big problems with this. First, the tetrad, and therefore the metric is buried inside the connection $\cA$, and can only be extracted upon identification of a subgroup $H$ that splits up the connection into $\mf{h}$ and $\mf{p}$ parts. The problem is that the Hodge dual operator $*$ is metric dependent. Specifically, it requires a metric and its associated volume form. So there is no natural, obvious way of expressing it in terms of the Cartan connection $\cA$. On the other hand, there is another sort of dual that acts on internal indices. This is the dual operator $\star=-i\g_5$, which exploits the fact the internal $Spin(3,1)$ vector space $\mb{V}$ is equipped with the Lorentzian metric $\eta_{IJ}$. Let's try to use this. By similarity with the Yang Mills action, let's try the dumbest thing possible:
\beq
S=\alpha \int_M \star \,F_\cA \, F_\cA\,.
\eeq
For simplicity let's focus on the de Sitter group (anti-de Sitter will be similar with a few minus signs here and there, but Poincar\'{e} requires different tricks). Recalling that there is an assumed trace over the Clifford algebra, and the trace of any odd number of gamma matrices is zero (and $Tr( \g_5 \g^I \g^J)=0$), we have 
\beq
S=\alpha \int_M \star\,R_\omega \, R_\omega - \frac{2}{\ell^2}\star e\,e\,R_\omega +\frac{1}{\ell^4} \star e\,e\,e\,e
\eeq
Pulling out the factor of $\frac{2}{\ell^2}$ and recalling $\ell^2=\frac{3}{\Lambda}$ we have
\beq
S=\alpha \int_M \star R_\omega \, R_\omega -\frac{2\alpha}{\ell^2}\int_M \star\,e\,e\,R_\omega -\frac{\Lambda}{6} \star \,e\,e\,e\,e\,.
\eeq
Now notice, provided we identify $\alpha=-\frac{3}{16 \pi G\Lambda}$, the last two terms are precisely the Einstein-Cartan action! Now what about the first term? Don't be fooled into thinking this term is a Yang-Mills term: $\star$ is not $*$. In fact the first term is a well-known topological term known as the Euler characteristic. Being topological, its variation does not effect the equations of motion. Thus, the simplest possible choice gives us precisely what we wanted (modulo topological terms). This is the Macdowell-Mansouri action:
\beq
S_{MM}=\alpha \int_M \star \,F_\cA \, F_\cA \,.
\eeq

One interesting fact: the coupling constant in front of the action, $\alpha$, is {\it dimensionless}, since it involves only the combination $G\Lambda$. This is tantalizing given that many of the simplest arguments for the non-renormalizability of perturbative quantum gravity follow from the simple fact that the coupling constant is {\it dimensionful}. But, I won't say anything more on this.

It appears at if we have now successfully written gravity as a type of gauge theory based not on the Lorentz group, but on the (A)dS group. But, we shouldn't be too quick to jump the gun. It is true that the action involves only the curvature of the (A)dS connection, but we need to check that the action is invariant under local $Spin(4,1)$ transformations. In fact, it is {\it not}. To see this, take a simple example. In the de Sitter case, consider a translation $g=\cos(\xi) {\bf{1}} +\sin(\xi) \, \h{V}_I \,\g_5 \g^I$ where $\h{V}_I$ is a spatial unit vector. Under this transformation $F_cA \rightarrow gF_\cA g^{-1}$ so 
\beqa
S_{MM}\rightarrow S'_{MM} &=& \alpha \int_M g^{-1} \star g \,F_\cA \, F_\cA \nn\\
&=& \alpha \int_M \cos(2\xi) \star F_{\cA} \, F_{\cA} +i \sin(2\xi) \,\h{V}_I \g^I \,F_\cA \, F_\cA\nn\\
&\stackrel{on-shell}{\approx}&  \alpha \int_M \cos(2\xi) \star F_\cA \, F_\cA 
\eeqa
So the action is not invariant under $Spin(4,1)$, not even on shell. It is invariant under $Spin(3,1)$, but this should be expected if it reproduces the Einstein-Cartan action. To use suggestive language, the symmetry of the gauge theory has been broken at the level of the action. In the next section I will discuss a model where the symmetry is retained in full, but broken spontaneously similar to the way that the Higgs breaks electroweak symmetry. 

\section{Breaking the symmetry: the Stelle-West model \label{SWModel}}
So, I have shown how to construct an action involving the {\it ingredients} of a(n) (A)dS gauge theory that falls just short of being a true gauge theory since it is not invariant under the full gauge group. This phenomenon should be familiar from the interactions of the standard model. For example, the $SU(2)\times U(1)$ symmetry of the electroweak interaction appears to be non-existent at low energies (more accurately, it is there, but it is realized in a complicated, non-obvious, non-linear way). The introduction of new fields to space out the V-A contact interactions showed that the interaction could be modeled using the ingredients of an $SU(2)\times U(1)$ gauge theory, but the true gauge theory invariant under this gauge group could not emerge until a model for spontaneously breaking the symmetry was constructed \cite{Weinberg:1967tq} (historically this all occurred in one step, but conceptually one can imagine it as a two step process). 

For the gauge formulation of gravity, Stelle and West constructed a simple model of symmetry breaking that is in close analogy with the Higgs mechanism \cite{Stelle:1979aj,Stelle:1979va,West:1978Lagrangian}\cite{Fukuyama-Ikeda}\cite{Fukuyama:1984}. Just as the full symmetry of the electroweak gauge theory is retained even in the so-called symmetry broken phase, so too is the $Spin(4,1)$ ($Spin(3,2)$) symmetry preserved in full in the Stelle-West extension of the Macdowell-Mansouri action. The symmetry is just realized in a non-obvious, non-linear way. One caveat should be mentioned. The Higgs model is a fully dynamic spontaneous symmetry breaking mechanism since the Higgs (presumably) lives a full life on its own having kinetic energy and propagating at will. On the other hand, as far as I am aware, nobody has given a fully dynamic extension of the Stelle-West model. The fields that break the symmetry are introduced purely in order to break the symmetry, and it is very difficult to construct physically realistic kinetic terms for the symmetry breakers. For this reason I will refer to the Stelle-West model as ``quasi-dynamic". This paragraph should scream {\it open problem}.

Let's dig deeper to try to find out why the Macdowell Mansouri construction does not retain the full $Spin(4,1)$ symmetry (for definiteness I will focus on the de Sitter group here, but not much changes in the anti-de Sitter case) but only that of the subgroup $Spin(3,1)$. The problem is the $\star$. This object is an non-dynamical matrix. If it were to transform like $\star\rightarrow g \star g^{-1}$, then the action would be invariant. However, we can't simply define a new transformation law and expect things to still make sense mathematically. But there is a simple way to make it work out naturally. 

First, notice that $\star$ is essentially the metric volume form of the $Spin(3,1)$ representation space. Being the $Spin(3,1)$ volume form (essentially $\epsilon_{IJKL}$), it is not invariant under $Spin(4,1)$. Can we construct the action using the $Spin(4,1)$ volume form? We can, but it requires the introduction of new ingredients. This is slightly easier to see in the adjoint representation, so let's try to build the action using $F^{AB}$. Any action of the form $\int F_{AB}\,F^{AB}$ is proportional to the second Chern-class and is therefore topological (and parity violating). We need to use the $SO(4,1)$ volume form $\epsilon_{ABCDE}$ somehow, but there are just not enough objects to saturate the indices:
\beq
S\stackrel{?}{=} \int_M (?)\epsilon_{ABCDE} F^{BC} \, F^{DE}\,.
\eeq
But the solution is now obvious: just introduce a new vector field $V^A$ living in the adjoint (vector) representation of $Spin(4,1)$. As such, it is a vector field living in a $5$-dimensional vector space. Now, suppose $V^A$ was a spatial vector with magnitude $V_A V^A=1$. Then we could always use the $Spin(4,1)$ symmetry to choose a gauge where $V^A=(0,0,0,0,1)$, or in other words $V^I=0$ but $V^4=1$. Then the object $V^A\epsilon_{ABCDE}$ in this gauge would look like $V^4 \epsilon_{4IJKL} \stackrel{*}{=}\epsilon_{IJKL}$ which is the $Spin(3,1)$ volume form. Of course this identification only holds in this gauge, but if the theory can be written in a gauge invariant way, the physics will still be the same regardless of the gauge. Use this to build the new action:
\beq
S= (?) \int_M  \epsilon_{ABCDE} \,V^A\,F^{BC} \, F^{DE} \ \bigg|_{V_A V^A=1}\,.
\eeq
At this stage, if the vector $V^A$ is restricted to live entirely on the hyperbola of magnitude $1$, then the $Spin(4,1)$ symmetry is realized in full, but in a nonlinear manner (just like the non-linear sigma model). The trick of Stelle and West is to lessen this constriction to some degree by implementing the constraint $V_A V^A$ quasi-dynamically, so that the symmetry of the full theory is realized linearly, but when the action is restricted on-shell it is realized non-linearly. The trick is simple, simply implement the constraint $(1-V_A V^A)=0$ via Lagrange multipliers. Try this:
\beq
S=(?) \int_M \epsilon_{ABCDE}\, V^A \,F^{BC}\, F^{DE} +\sigma (1-V_A V^A) 
\eeq
where $\sigma$ is an arbitrary four-form serving as a Lagrange multiplier. Clearly varying the action with respect to $\sigma$ implements the constraint we wanted. But things are less trivial than they may seem -- we still have to show that the constraint is compatible with the full set of equations of motion, which can be checked by varying with respect to $V^A$. In fact, the two on-shell surfaces are compatible. Varying with respect to $V^A$ simply imposes conditions completely constraining the Lagrange multiplier $\sigma$, and nothing more. Specifically, in the right gauge, the Lagrange multiplier becomes proportional to (the dual of) one of the Kretschmann scalars, $\sigma \sim \epsilon_{IJKL}W^{IJ}\w W^{KL}$, where $W^{IJ}$ is the Weyl tensor (we've already used $C^{IJ}$), which is a free parameter in vacuum general relativity. So the constraint is compatible with the equations of motion. Work through the details and you will find that this is actually a delicate balance -- had we chose the constraint to be $(1-V_A V^A)^2$ as opposed to $(1-V_A V^A)$ it would not have worked because varying with respect to $V^A$ would have put constraints on the Kretschmann scalar that are not required by the Einstein equations.

One further comment is in order. One may object to this ever being a physically realistic model since the field that breaks the symmetry, being an object living in a five-dimensional vector space, has no precedent in nature and appears unlikely to be physical. Speculation about the physicality of a field that naturally takes values in an internal vector space aside, I should mention that this is a bare model where the order parameter that breaks the symmetry has been isolated. More specifically, generically in order to break the symmetry of a gauge group $G$ to a subgroup $H$, an order parameter living in the coset space $G/H$ must freeze out. But, this field is simply  the order parameter, and nothing prevents it from being a composite object, or even an emergent field arising from extremely complicated dynamics. For example, in the context of the de Sitter group, one can combine the vector current $J^I=\bar{\psi}\g^I \psi$ with the pseudo-scalar $\rho=\bar{\psi} \g_5 \psi$ into a single current $V^A=(J^I, \rho)$ which one can check transforms like an ordinary $SO(4,1)$ vector field. In this context, the order parameter emerges upon a type of fermion condensation that fixes the magnitude of the vector field $V_A V^A=1$. For more details on a model of this form see \cite{Randono:Condensate}. Some work, but not enough has been done on coupling matter to the gauge framework of gravity \cite{Fukuyama-Ikeda}. For point like defects with internal spin degrees of freedom, the matter action can be described by coupling the theory to a Wilson line of the connection \cite{Freidel:2006hv}\cite{Fairbairn:2008hy}.

\subsection{Physical quantities in geometric vs topological gauge \label{PhysicalQuant}}
So I have now shown how the $Spin(4,1)$ symmetry can be retained while still reproducing all the features of General Relativity. The gauge freedom of the theory has been enlarged from $Spin(3,1)\rtimes Diff(M)$ to $Spin(4,1)\rtimes Diff(M)$. This gives another level of gauge freedom that can sometimes make the invariant physical content even more difficult to see. This doesn't sound good, but hopefully I will convince you later that we can get new physics from this framework. For now I want to give an extreme example illustrating how the same physical quantity can take on an entirely different character in two different $Spin(4,1)$ gauges. 

To make the example as clear as possible, let's temporarily consider the Euclidean analog of our (A)dS gauge theory. Everything proceeds exactly the same way, the only difference being that the gauge group is $G=Spin(5)$, and the stabilizer subgroup is $H=Spin(4)$. The homogenous Klein geometry serving as our model space is $Spin(5)/Spin(4)$ which is just the sphere $\mb{S}^4$ equipped with the ordinary constant curvature, zero torsion geometry. This is our model geometry. So suppose that the manifold $M$ is topologically $\mb{S}^4$ and the geometry itself (described by the connection $\cA=\omega+\frac{1}{\ell} e$) is actually this constant curvature, zero torsion geometry (of radius $\ell$). Consider in this case the following two integrals (in this section, and this section alone, $A,B,C,..=\{1,2,3,4,5\}$ and $I,J,K,...=\{1,2,3,4\}$):
\beq
\frac{1}{\frac{1}{2}\pi^2 \ell^4} \int_M \frac{1}{4!} \epsilon_{IJKL} \,e^I \, e^J \, e^K \, e^L \quad\quad \frac{1}{12\pi^2} \int_M  \epsilon_{ABCDE} \,\hat{V}^A \,d\h{V}^B \, d\h{V}^C \, d\h{V}^D \, d\h{V}^E\,. \label{Volumes}
\eeq
Here $\hat{V}^A$ is just the order parameter I introduced in the last section, just already constrained by $\h{V}_A \h{V}^A=1$, which is why I gave it a hat. The integral on the left should be familiar -- it is just the volume of the four-sphere, normalized so that the integral equals $1$ (the volume of the four-sphere of radius $\ell$ is $\frac{1}{2} \pi^2 \ell^4$).

The integral on the right is likely less familiar (unless you are a topologist). Notice, it does not involve any connection, just the exterior derivative. In fact, it is a topological integral \cite{Rajaraman:Instantons}\cite{Kobayashi}. Recall that the constraint $\h{V}_A\h{V}^A=1$ defines the four sphere embedded in the vector space $\mb{R}^5$. As such, the costrained vector field $\h{V}^A=\h{V}^A(x)$ can be viewed as a map $\h{V}: M\simeq \mb{S}^4 \rightarrow \mb{S}^4$. From a topological perspective, these maps fall into a set of discrete classes labelled by what is referred to as the winding number of one 4-sphere onto the other. The winding number is an integer, and when there is a clear sense of composition of maps (which will not be important in this section, but will become clear later), they form a group. Specifically the group in question is denoted $\pi_4(\mb{S}^4)$, and it is a well known result in topology that this is just the group of integers $\mb{Z}$ (more generally $\pi_n(\mb{S}^n)=\mb{Z}$). Thus, the integral on the right is an integer, and furthermore, being topological it is invariant under small deformations of $\h{V}^A$.

In fact, as you may have guessed the two integral are the same physical quantity of the $Spin(5)$ gauge theory, just written in two different gauges. To see this, let's first see how we can identify the tetrad in a $Spin(5)$ invariant way. To do this, start in a gauge where $\h{V}^A=(0,0,0,0,1)$. Then from the previous section it should be clear that $\frac{1}{\ell}e^I=\cA^I{}_5$. This can be written in a more covariant way by noting in this guage $D_\cA \h{V}^I=\cA^I{}_5$, and $D_\cA \h{V}^5=0$. In an arbitrary gauge, $D_\cA \h{V}^A$ has only components perpendicular to $\h{V}^A$ by which I mean that $D_\cA \h{V}^A=\pi^A{}_B \,D_\cA \h{V}^B$ where $\pi^A{}_B=\delta^A{}_B-\h{V}^A \h{V}_B $ is the perpendicular projector of $\h{V}^A$ (i.e. $\pi^A{}_B \h{V}^B=0$). So, the most covariant way to define an object with all and only the information defined in the tetrad is by the object $\ell D_\cA \h{V}^A$. The metric emerges as
\beq
\bm{g}=\ell^2\,\delta_{AB} \,D_{\cA}\h{V}^A \otimes D_{\cA}\h{V}^B \,.
\eeq
Now, the normalized volume can be written 
\beq
\frac{V(M)}{\frac{1}{2}\pi^2 \ell^4}= \frac{1}{\frac{1}{2}\pi^2 \ell^4} \int_M \frac{\ell^4}{4!}\, \epsilon_{ABCDE} \,\hat{V}^A \,D_{\cA} \h{V}^B \, D_{\cA}\h{V}^C \, D_{\cA}\h{V}^D \, D_{\cA}\h{V}^E
\eeq
as can easily be checked by noting that in the gauge where $V=(0,0,0,0,1)$ and $\ell \,D_{\cA}\h{V}^I =e^I$, the integral reduces to the volume integral on the left in (\ref{Volumes}). 

How can one show that this is equal to the integral on the right in (\ref{Volumes})? Clearly it has roughly the same form. The trick is to somehow get rid of the gauge potential $\cA$. In fact we can actually do this by a gauge transformation. To see this, recall from the previous discussion that the homogenous Klein geometry as an actual geometry of the gauge theory is the (unique in our case) non-degenerate geometry on the fixed topology $M\simeq \mb{S}^4$ satisfying $F_{\cA}^{AB}=0$. So the $Spin(5)$ connection is flat. Now there is a special property of flat connections: if the manifold is in some sense topologically trivial, which in this case means that all loops embedded in the manifold can be smoothly contracted to a single point (more mathematically, $\pi_1(M)=0$, which is true of $\mb{S}^4$) then all flat connections are equivalent modulo gauge transformations. This means that any flat connection can be obtained from any other by a gauge transformation. There is no simpler flat connection than the trivial connection $\cA=0$. But the previous statement means that all flat connections can be obtained by gauge transforming this trivial connection, or vice-versa: all flat connections can be gauge transformed to the trivial connection. Thus, we can find a gauge where $\cA=0$, and in this gauge the normalized volume is clearly
\beq
\frac{V(M)}{\frac{1}{2}\pi^2 \ell^4}=\frac{1}{\frac{1}{2}\pi^2} \int_M  \frac{1}{4!}\epsilon_{ABCDE} \,\hat{V}^A \,d\h{V}^B \, d\h{V}^C \, d\h{V}^D \, d\h{V}^E
\eeq 
which is the right hand side of (\ref{Volumes}), thereby relating the geometric (left) to the topological (right) integral.

I'd like to stress this last sentence. This is one of the peculiar features of gravity as a gauge theory, which will eventually allow us to get new physics -- there are both geometric and topological phases of the theory. Sometimes the two phases are really related by a shift in perspective as was the case here, but sometimes they are not.

One more comment before I move on -- this section was written in the Euclidean sector because the analogous integrals in the Lorentzian sector are divergent. But, as we will see shortly, there are very similar integrals in the Lorentzian sector that are not divergent, and it turns out they distinguish a class of ground states of the theory.

%%%%%%%%%%%%%%%%%%%%%%%%%%%%%
%%%%%%%%%%%%%%%%%%%%%%%%%%%%%

\section{de Sitter gauge theory \label{dSGaugeTheory}}
In the last few sections I presented a new formalism for gravity that reveals an underlying local de Sitter, anti-de Sitter, or Poincar\'{e} symmetry, that is spontaneously broken in the theory. Let's now try to go beyond the formalism to see if we can get new physics out of the theory.

Look at the Einstein-Cartan field equations again, written in the notation of differential forms (here written in vacuum):
\beqa
\epsilon_{IJKL}\,e^J \w \left( R_\omega{}^{KL}-\frac{\Lambda}{3} e^K \w e^L \right)&=& 0 \label{EC1b} \\
\epsilon_{IJKL}\,D_{\omega} (e^K \w e^L) &=& 0 \label{EC2b}\,.
\eeqa

Buried in there is something new that is not present in the ordinary form of the Einstein equations $G_{\mu\nu}=0$ and $T^\alpha{}_{\mu\nu}=0$. The Einstein-Hilbert field equations involve not just the metric $g_{\mu\nu}$ but also its inverse $g^{\mu\nu}$. On the other hand, (\ref{EC1b}) and (\ref{EC2b}) only involve the coframe $e^I$ and make no reference to its dual, the frame $\bar{e}_I$. What this means is that whereas the Einstein-Hilbert field equations only make sense when the metric is non-degenerate (invertible), the Einstein-Cartan  field equations make perfect sense even when the tetrad (or metric) is degenerate. This allows for a whole new set of solutions that don't exist in ordinary general relativity. There is no additional field equation that says the determinant of the tetrad must be non-zero (see \cite{Mielke:1992gk} for an attempt at avoiding this problem, and \cite{Bengtsson:1997wr} for some solutions to the Einstein-Cartan equations involving degenerate metrics). Moreover, this would be difficult to naturally implement by a variational principle since such an condition is more of a {\it non}-equation\footnote{Thanks to Derek Wise for this terminology.} than an equation, being equivalent to $\epsilon_{IJKL}e^I \w e^J \w e^K \w e^L \neq 0$.  

Take a simple (almost trivial) example. Suppose I took $e^I =0 $ and $\omega^{IJ}=0$. Nothing stops me from doing this (provided it is allowed by the topology) -- this is a perfectly good set of data: it is as smooth as it gets and there are no pathologies to the field. Of course, this is a solution to the Einstein-Cartan equations of motion. From the geometric perspective, this is ludicrous. It does not define a geometry. But, this is an article on gravity as a gauge theory. From the perspective of gravity as a gauge theory this configuration, being part of the (A)dS connection $\cA$, is perfectly natural. It is more than natural -- we have already seen (in the context of the Euclidean theory but the same holds in the Lorentzian case) that the connection defining the constant curvature zero torsion geometry is gauge related to the trivial $\cA=0$ connection.

So now a lengthy debate could ensue about whether these degenerate configurations should be considered. Rather than pursue this debate, I will run with it. Let's see what new things we can get when gravity is viewed as a gauge theory.

\subsection{Why the de Sitter group? \label{WhydS}}
For the rest of this article I will focus on the de Sitter ($Spin(4,1)$) gauge theory. {\it Why de Sitter?} There are two reasons. First, the best interpretation of cosmological observations shows that the cosmological constant is non-zero, and it is positive \cite{WMAP}. In a field where observational evidence is scant, the limited data that we do have should be taken seriously. Second, from a purely theoretical perspective, the de Sitter gauge theory has very interesting properties that are not shared by the anti-de Sitter or Poincar\'{e} theories. Exploring these properties will be the focus of the next few sections. 

\subsection{Topological aspects of de Sitter space and the de Sitter group \label{GroupTopology}}
de Sitter space and the de Sitter group has a rich topological structure that allows for physics that cannot occur in the anti-de Sitter or Poincar\'{e} case. There are two topological aspects of the de Sitter gauge theory that are relevant for this discussion. First, there is the topology of the homogenous Klein geometry $G/H$. This model geometry will serve as a ``ground state" (we are using this term loosely since as I will show there are good indications that it is not a stable ground state, hence the scare quotes). In the Poincar\'{e} case, $G/H=\overline{ISO}(3,1)/Spin(3,1)=M_4 \simeq \mb{R}^{3,1}$, which is a rather dull topology. Similarly, for anti-de Sitter $Spin(3,2)/Spin(3,1)\simeq \mb{S}\times \mb{R}^3$, which is slightly more interesting, but when the universal cover is taken to remove the closed timelike curves, the topology becomes again $AdS_4\simeq \mb{R}^4$. For de Sitter, the Klein geometry is $Spin(4,1)/Spin(3,1)\simeq \mb{R} \times \mb{S}^3$, which as it turns out does allow for very interesting phenomena. 

Second, there is the topology of the gauge group itself. All the gauge groups we are interested in are non-compact. One nice feature of non-compact semi-simple, simply connected Lie groups is that the non-compactness is contractible in the topological sense. What this means is that the Lie group itself, $G$, is homeomorphic (which implies they have the same topology) to $G\approx G_0\times \mb{R}^n$ where $H$ is referred to as a maximally compact subgroup \cite{HofmannMorris:Groups}\cite{Kobayashi}\cite{Helgason:Differential}. The maximal compact subgroup is essentially unique, meaning it is unique up to conjugation ($G_0\rightarrow gG_0g^{-1}$ for $g\in G$). So, all the interesting topological properties are contained in the maximal compact subgroup $G_0$. For example, for our stabilizer subgroup $H=Spin(3,1)$, the maximal compact subgroup is $Spin(3)=SU(2)$. Thus, $Spin(3,1)\approx SU(2)\times \mb{R}^3$. The $SU(2)$ can be identified (up to conjugation) with the rotation subgroup, and the remaining part is formed by the set of boosts. In fact there is a very general statement about spin groups that the maximal compact subgroup is \footnote{The quotient in this relation means the following: given a pair $(g_1,g_2)\in Spin(p)\times Spin(q)$, the quotient means that this is equivalent to $(-g_1,-g_2)$. This can be thought of as a statement about the identity and its negative. The subgroups $Spin(p)$ and $Spin(q)$ share the identity element $(1,1)$ so the negative of it is $(-1,-1)$. This means $(-g_1,-g_2)=(-1,-1) \times (-1,-1)\times (g_1,g_2)=(g_1,g_2)$.} $G_0=Spin(p)\times Spin(q)/\{\{1,1\},\{-1,-1\}\}$. This allows us to build the table
\beqa
\begin{array}{ccccc}
\overline{ISO}(3,1) &\approx & SU(2) \times \mb{R}^{7} &\simeq& \mb{S}^3 \times \mb{R}^{7} \nn\\
Spin(3,2) & \approx& SU(2) \times U(1) \times \mb{R}^6 & \simeq& \mb{S}^3 \times \mb{S}^1 \times \mb{R}^6 \nn\\
Spin(4,1)  &\approx& SU(2) \times SU(2) \times \mb{R}^4 &\simeq &\mb{S}^3 \times \mb{S}^3 \times \mb{R}^4 \,.
\end{array}
\eeqa
In both the Poincar\'{e} and the anti-de Sitter case, the $SU(2)$ part comes from the maximal compact subgroup of the stabilizer group $H=Spin(3,1)$. This will be important. In the de Sitter case, the maximal compact subgroup is $Spin(4)=SU(2)\times SU(2)$. To understand this, recall that the group $Spin(4,1)$ comes from double-covering the isometry group of de Sitter space, which is $SO(4,1)$. The topology of the space is $\mb{R}\times \mb{S}^3$ and the $\mb{R}$ is just the time direction. So suppose we considered the set of isometries that don't do anything to the time axis (i.e. time independent isometries). These are just the isometries of the three sphere, which forms the group $SO(4)$. Imagine you are sitting at some fixed point in a three sphere (not de Sitter, which is slightly more complicated, just a three sphere). With a powerful enough telescope, anywhere you look if you look far enough you will see the back of your head. You can turn your head in any direction and the space will look the same. This forms the rotation subgroup giving an $SO(3)$ subgroup. But, you can also walk (translate) in any direction. But since the space is compact, walk far enough and you will end up where you started. In total, the set of things you can do (move your head and walk) forms the group $SO(4)=SO(3)\times SO(3)$, the double cover of which is $Spin(4)=SU(2)\times SU(2)$. They key point is that the spatial translations are also part of the compact group, since you can only move so far before ending up where you started.

Here is the most interesting fact about the de Sitter group (for our purposes). The maximal compact subgroup\footnote{Note this is also true of the anti-de Sitter case since the $U(1)$ in $SU(2)\times U(1)$ does not come from the Lorentz subgroup (it comes from the compact set of time translations of the homogenous space prior to taking the universal cover) but for our purposes, the $U(1)$ is not as interesting.} does not live in the stabilizer group: $G_0=Spin(4) \nsubseteq H=Spin(3,1)$. As we will see this allows for the construction of an infinite class of ``ground states" of the de Sitter gauge theory.

\subsection{Winding numbers of $SU(2)$ \label{WindingNumbers}}
From the previous section, it was clear that the most interesting topological features of the gauge group come from $SU(2)$ subgroups. So let me now make a short digression to explain some basic properties about $SU(2)$ and $\mb{S}^3$ and maps between the two.

First, we recall that topologically the group is $SU(2)\simeq \mb{S}^3$. The easiest way to see this is to write the generic group element (in this section, hatted indices $\h{a},\h{b},\h{c},...=\{1,2,3,4\}$ and $\h{i},\h{j},\h{k},...=\{1,2,3\}$)
\beq
SU(2) \ni  g=X^4 \,\bm{1} +X^{\h{i}} \,i \sigma_{\h{i}}
\eeq
where $\sigma^{\h{i}}$ are the ordinary Pauli matrices. The condition $g^\dagger =g^{-1}$ implies
\beq
\delta_{\h{a}\h{b}} X^{\h{a}} X^{\h{b}}=1 \label{g}
\eeq
which defines the three sphere embedded in $\mb{R}^4$. Now suppose I gave you a manifold $M$ which itself was topologically $\mb{S}^3$ and defined a group element at each point $g=g(x)$ which is continuos and differentiable. The group field $g(x)$ can then be though of as a map $g:M\rightarrow SU(2)$ but since $SU(2) \simeq \mb{S}^3$ and $M\simeq \mb{S}^3$, it can be though of as a map $g:\mb{S}^3 \rightarrow \mb{S}^3$. Of course at this point, the map need not be $1$-to-$1$ or onto. In fact, we can classify such maps by discrete classes distinguished by topological features of the map. Two maps are said to be in the same class if there is a sequence of smooth infinitesimal deformations that takes one map to the other. So, any map that can be obtained by smoothly deforming the identity map, is equivalent under this equivalence relation to the identity. This is the trivial sector, referred to as the group of {\it small} gauge transformations (if we are regarding $g(x)$ as defining a gauge transformation of some field) denoted $SU(2)_0$. Now suppose instead we had a map $h(x)$ such that $h:\mb{S}^3\rightarrow \mb{S}^3$ is one-to-one and onto. For example, we could take
\beq
g\equiv \overset{1}{g} = Y^4 \,\bm{1} +Y^{\h{i}} \,i \sigma_{\h{i}}
\eeq
with
\beqa
Y^{\h{4}}=\cos{\chi} \quad \quad Y^{\h{i}}=\sin{\chi} \,\wt{Y}^{\h{i}}  \quad \quad \h{i}=\{\h{1},\h{2},\h{3}\} 
\eeqa
and
\beqa
\wt{Y}^{\h{1}}&=&\sin{\theta} \cos{\phi} \nn\\
\wt{Y}^{\h{2}}&=&\sin{\theta} \sin{\phi} \nn\\
\wt{Y}^{\h{3}}&=&\cos{\theta}\,. \label{Y}
\eeqa
This defines a smooth one-to-one map from $\mb{S}^3$ onto $SU(2)$. There is no way to smoothly deform this map to the identity since the map must remain onto under small deformations, but the identity map $g(x)=\bm{1}$ sends every point on $\mb{S}^3$ to one point on $SU(2)$. So this map falls in a different sector. If the identity map is the $n=0$ sector call this the $n=1$ sector. If $\overset{1}{g}$ is one-to-one and onto, then $\overset{2}{g}\equiv \overset{1}{g}{}^2$ is {\it two-to-one} since if $\overset{1}{g}(x_1)=-\overset{1}{g}(x_2)$, then $\overset{2}{g}(x_1)=\overset{2}{g}(x_2)$ and {\it onto} since for every group element $b$ there is an $a$ such that $b=aa$. This means that $\overset{1}{g}$ can be used as the generator of an equivalence class of maps labelled by an integer $n$, whose typical elements are $\overset{n}{g}=\overset{1}{g}{}^n$. Each map $\overset{n}{g}$ (for $n\neq 0$) is an $|n|$-to-$1$ map from $M\simeq \mb{S}^3$ onto $SU(2)\simeq \mb{S}^3$.

It is easy to see that under the equivalence relation (call it $\approx$), the $n$-sectors form an additive abelian group since $\overset{m}{g}\approx \overset{1}{g}{}^{m}$ means $\overset{m}{g} \overset{n}{g} \approx \overset{1}{g}{}^{m+n} \approx \overset{n}{g} \overset{m}{g}$. Clearly, the abelian group is equivalent to the additive group of integers, $\mb{Z}$. The integer labelling the sector in which the group element lives is often referred to as the winding number of the map.

More generally, the set of maps from $\mb{S}^m$ to some manifold $X$ modulo the equivalence relation defined by homeomorphisms (small deformations), forms a group $\pi_m(X)$. For our case $m=3$ and $X=SU(2)\simeq \mb{S}^3$, so $\pi_3 (\mb{S}^3) =\mb{Z}$.

Now suppose I have some map $g:\mb{S}^3\rightarrow SU(2)$. Is there an easy way to determine what sector it lives in? In fact there is a generic integral that one can write down that will give the winding number. The integral is the following (written in the two-dimensional Pauli matrix representation):
\beq
W(g)=\frac{1}{24\pi^2} \int_{\mb{S}^3} Tr\left( dg\,g^{-1} \w dg\,g^{-1} \w dg\,g^{-1} \right)\,.
\eeq
This integral is topological and therefore invariant under small deformations of the field $g(x)$ as can be checked by computing the change in the integral under $g\rightarrow g+\dl g$. To see that it does give the winding number, for the group element denoted by (\ref{g}) the following identity holds (this takes some algebra, which I won't show)
\beq
\frac{1}{24\pi^2} \int_{\mb{S}^3} Tr\left( dg\,g^{-1} \w dg\,g^{-1} \w dg\,g^{-1} \right)=\frac{1}{12\pi^2}\int_{\mb{S}^3} \epsilon_{\h{a}\h{b}\h{c}\h{d}} \,X^{\h{a}} dX^{\h{b}}\,dX^{\h{c}}\,dX^{\h{d}}\,.
\eeq
This last integral should be slightly familiar from section \ref{PhysicalQuant} where we saw the same expression in the five-dimensional case. Just as there, this integral gives the winding number of $X^{\h{a}}$ viewed as a map $X:\mb{S}^3 \rightarrow \mb{S}^3$ (recall $X^{\h{a}}$ is constrained by $X_{\h{a}}X^{\h{a}}=1$ to live on a three-sphere). 

This integral is topological, so it is invariant under small deformations of $X^{\h{a}}$. So it doesn't really matter too much what specific form we use for $X^{\h{a}}$, the result will only depend on topological information about the map. Let me first compute it for the group element $\overset{1}{g}$ where $X^{\h{a}}=Y^{\h{a}}$ given above. From there it will be a simple matter to extend it to $\overset{n}{g}$. So, suppose at first that $Y^{\h{a}}$ were not constrained to live on the sphere but was just an arbitrary vector field, and further suppose that the manifold $\Sigma\simeq \mb{S}^3$ was embedded in the usual way in $\mb{R}^4$. Then we could write down the volume form $\wt{\sigma}_4=\frac{1}{4!}\epsilon_{\h{a}\h{b}\h{c}\h{d}} \,dY^{\h{a}} \w dY^{\h{b}}\w dY^{\h{c}}\w dY^{\h{d}}$. We can think of $Y^{\h{a}}$ as the Cartesian coordinates on $\mb{R}^4$. As such, we can always change coordinates -- let's choose spherical coordinates. In spherical coordinates the volume form is $\wt{\sigma}_4=r^3 \sin^2\chi \,\sin\theta \,dr \w d\chi \w d\theta \w d\phi$. Now the trick is to restrict this integral to the three sphere by imposing the constraint $Y_{\h{a}}Y^{\h{a}}=1$. We first note that given the radial vector field $r \frac{\p}{\p r}=Y^{\h{a}} \frac{\p}{\p Y^{\h{a}}}$ we have $\wt{\sigma}_4(Y^{\h{a}} \frac{\p}{\p Y^{\h{a}}})=\frac{1}{3!} \epsilon_{\h{a}\h{b}\h{c}\h{d}} \,Y^{\h{a}} dY^{\h{b}}\,dY^{\h{c}}\,dY^{\h{d}}$. In spherical coordinates, this is $\wt{\sigma}_4(Y^{\h{a}} \frac{\p}{\p Y^{\h{a}}})=r^4 \sin^2\chi \,\sin \theta\,d\chi\w d\theta \w d\phi$. In these coordinates it is trivial to impose the constraint $Y_{\h{a}}Y^{\h{a}}=1$: we just set $r=1$. So in total, the integral becomes
\beqa
\frac{1}{2\pi^2}\int_{\mb{S}^3} \frac{1}{3!} \epsilon_{\h{a}\h{b}\h{c}\h{d}} \,Y^{\h{a}} dY^{\h{b}}\,dY^{\h{c}}\,dY^{\h{d}} &=& \frac{1}{2\pi^2}\int_{\mb{S}^3} \wt{\sigma}_4(r\frac{\p}{\p r}) \Big|_{r=1}\nn\\
 &=& \frac{1}{2\pi^2}\int^{\pi}_{\chi=0}\int_{\theta=0}^\pi \int_{\phi=0}^{2\pi}  \sin^2\chi \,\sin\theta \, d\chi \w d\theta \w d\phi \nn\\
 &=& 1\,.
\eeqa
To extend the integral to $\overset{n}{g}$, we first note that it is a simple matter to show that given a group element $g=g_1g_2$, the winding number satisfies $W(g=g_1g_2)=W(g_1)+W(g_2)$ which is a reflection of the fact that $\pi_3(\mb{S}^3)=\mb{Z}$ is an abelian group. Thus 
\beq
W(\overset{n}{g}=\overset{1}{g}{}^n)=n\,W(\overset{1}{g})=n\,.
\eeq

For future reference, let me point out that the group element $\overset{n}{g}$ takes the simple form
\beq
\overset{n}{g}=\overset{n}{X}{}^{\h{4}}\,\bm{1}+\overset{n}{X}{}^{\h{i}} \,i\sigma_{\h{i}}
\eeq
with 
\beqa
\overset{n}{X}{}^{\h{4}}=\cos(n\chi) \quad \quad \overset{n}{X}{}^{\h{i}}=\sin(n\chi) \,\wt{Y}^{\h{i}}  \quad \quad \h{i}=\{\h{1},\h{2},\h{3}\} 
\eeqa
and $\wt{Y}^{\h{a}}$ defined as in (\ref{Y}).

\subsection{A case study: exotic geometries on the three-sphere \label{3Sphere}}
Having finished our digression into the winding numbers of $SU(2)$ we are now ready to use these tools. The ultimate goal is to employ them in the de Sitter gauge theories to construct exotic geometries. But for simplicity I will begin with the three-sphere example. I'm going to use the group elements given above to construct new geometries on the three-sphere. 

Recall the relevant group here is $Spin(4)=SU(2)\times SU(2)$, and the stabilizer subgroup is the group $H=SU(2)_{diag}$ consisting of diagonal elements $(g,g)$. The ordinary three-sphere geometry is then obtained by the homogenous Klein geometry $Spin(4)/SU(2)_{diag}$. 

I am going to adopt some new notation to simplify some calculations. Much of this notation will carry over to the de Sitter case. First, to distinguish the two copies of $SU(2)$, I will label them by $SU(2)_{\ua}$ and $SU(2)_{\da}$. The two generators of $\mf{su}(2)_{\ua}$ and $\mf{su}(2)_{\da}$ are denoted by
\beq
\tau^i_{\ua}=\left[\begin{matrix} \frac{i}{2} \sigma^i & 0 \\ 0 & 0\end{matrix} \right] \quad \quad \tau^i_{\da}=\left[\begin{matrix} 0 & 0 \\ 0 &  \frac{i}{2} \sigma^i \end{matrix} \right]\,.
\eeq
The corresponding connection is written $A=A_{\ua}+A_{\da}$. The Cartan decomposition that yields the ordinary constant curvature geometry on the three sphere is the diagonal decomposition, so we denote
\beq
\tau^i=\left[\begin{matrix} \frac{i}{2} \sigma^i & 0 \\ 0 & \frac{i}{2} \sigma^i\end{matrix} \right] \quad \quad \eta^i=\left[\begin{matrix} \frac{i}{2} \sigma^i & 0 \\ 0 &  -\frac{i}{2} \sigma^i \end{matrix} \right]
\eeq
where $\tau^i$ forms the Lie algebra $\mf{su}(2)_{diag}$ and the complement spanned by $\eta^i$ forms the set of transvections. Corresponding to this decomposition the connection is written $A=w+\frac{1}{\ell} E=w^i\tau_i +\frac{1}{\ell} E^i \eta_i$ or
\beq
A =\left[ \begin{matrix} A^i_{\ua}\, \frac{i}{2}\sigma_i  & 0 \\ 0 & A^i_{\da}\, \frac{i}{2}\sigma_i   \end{matrix} \right]
=\left[ \begin{matrix} (w^i+\frac{1}{\ell}E^i)\, \frac{i}{2}\sigma_i & 0 \\ 0 & (w^i-\frac{1}{\ell}E^i)\, \frac{i}{2}\sigma_i\end{matrix} \right] \,.
\eeq
In this formula, $E^i$ is the ordinary triad on the three sphere and $w^i{}_j=\epsilon^i{}_{jk}w^k$ is the ordinary three-dimensional spin connection in the adjoint representation of $SU(2)$. 

Now, recall that the decomposition $Spin(4)=SU(2)\times SU(2)$ means that topologically $Spin(4)\simeq \mb{S}^3 \times \mb{S}^3$. If we take our manifold to also have topology $\Sigma\simeq \mb{S}^3$, this means that a gauge transformation coming from $h(x)$ can be though of as a map $h:\mb{S}^3 \rightarrow \mb{S}^3 \times \mb{S}^3$. The homotopically distinct maps of this form fall into the discrete classes that form the group (under composition of maps) $\pi_3(\mb{S}^3 \times \mb{S}^3)=\mb{Z}\oplus \mb{Z}$. Thus any such map $h$ can be labelled by two integers (the winding numbers) denoting the sector in which the map lives. We already know how to construct these. Just use what we know from $SU(2)$ and apply it to $SU(2)_{\ua}\times SU(2)_{\da}$. With $Y^{\h{a}}$ defined as in (\ref{Y}), write
\beq
\ou{1}{h}{0} =\left[\begin{matrix} X_{\h{4}}\bm{1} + X_{\h{i}} \,i\sigma^i & 0 \\ 0 & \bm{1} \end{matrix} \right] \quad \quad \ou{0}{h}{1} =\left[\begin{matrix} \bm{1} & 0 \\ 0 & X_{\h{4}}\bm{1} + X_{\h{i}} \,i\sigma^i  \end{matrix} \right]\,.
\eeq
Now, define
\beq
\ou{m}{h}{n}\equiv \ou{1}{h}{0}{}^m\,\ou{0}{h}{1}{}^n\,.
\eeq
The winding number of this group element is $W(\ou{m}{h}{n})=W(\ou{1}{h}{0}{}^m)+W(\ou{0}{h}{1}{}^n)=m+n$.

With these tools available, let's try to construct geometries. First let's try to get the ordinary constant-curvature three-sphere geometry. The first thing to notice is that the curvature splits as usual into
\beq
F_A = \underbrace{R_\omega +\frac{1}{\ell^2} E\w E}_{\mathfrak{su}(2)_{diag} \text{-valued}} \ \oplus\  \underbrace{\frac{1}{\ell} D_\omega E}_{\mathfrak{p}\text{-valued}}
\eeq
and the constant curvature, zero-torsion geometry satisfies $F_A=0$. Thus, the homogenous Klein geometry is described by a flat connection. Recall that all flat connections on the three-sphere are related by a gauge transformation (since $\pi_1(\mb{S}^3)=0$). This means that I can always relate any flat connection to the trivial connection $A=0$ by a gauge transformation so that any flat connection satisfies
\beq
A=-dh \,h^{-1}
\eeq
for some $h\in Spin(4)_\Sigma$. The trick is then to choose the right $h$ such that when we extract the triad from the Cartan decomposition $A=w+\frac{1}{\ell}E$, it gives us the triad we wanted. Try this:
\beq
\ou{0}{A}{1}\equiv\ou{0}{w}{1}+\frac{1}{\ell}\ou{0}{E}{1}=-d \ou{0}{h}{1} \, \ou{0}{h}{1}{}^{-1}\,.
\eeq
It takes a bit of work, but the triad can be extracted to give
\beq
\ou{0}{E}{1}{}^i=\ell \left( Y^{\h{4}} \,d Y^{i} -Y^{i} \,dY^{\h{4}} -\epsilon^i{}_{jk} \,Y^{j} \,dY^k \right)\,.
\eeq
I don't expect many people will recognize this triad, but it plays an important role. The three sphere is special in that it is one of a distinguished class of spheres (including only $\mb{S}^0$, $\mb{S}^1$, $\mb{S}^3$, and $\mb{S}^7$) that is {\it parallelizable}. This means that one can find a set of three linearly independent vector fields (or one-forms) that are defined {\it globally}. And, the triad above is one of these globally defined sets. Although the coordinates $\{\chi,\theta,\phi\}$ fail at the poles, the triad is perfectly well defined there. Take the pole at $\chi=0$. The triad there is $\ou{0}{E}{1}{}^i=\ell dY^i$ which is perfectly well defined if we interpret, for example, the $Y^i$ as Cartesian coordinates projected onto the sphere. But, what geometry does this define? An explicit calculation show that the metric is
\beq
\ou{0}{\bm{g}}{1}=\delta_{ij} \,\ou{0}{E}{1}{}^i \otimes \ou{0}{E}{1}{}^j= \ell^2 \delta_{\h{a} \h{b}} \,dY^{\h{a}} \otimes dY^{\h{b}}\,. 
\eeq
Recalling we have the constraint $Y_{\h{a}} Y^{\h{a}}=1$, the last term shows that the metric is the ordinary flat Euclidean metric on $\mb{R}^4$ restricted to the three-sphere. Of course we know this metric, it is just
\beq
\ou{0}{\bm{g}}{1}= \ell^2 \left(d\chi^2 + \sin^2\chi \left( d\theta^2 +\sin^2\theta \,d\phi^2 \right) \,.\right) 
\eeq
So, we now know how to relate the constant curvature geometry, zero-torsion geometry defined as a $Spin(4)$ Cartan connection $\ou{0}{A}{1}$ on $\mb{S}^3$ to the trivial connection $A=0$. We just gauge transform by $\ou{0}{h}{1}$. Can we keep going with this? What if instead we transformed by $\ou{0}{h}{1}{}^2=\ou{0}{h}{2}$? Or more generally by $\ou{m}{h}{n}$? Define
\beq
\ou{m}{A}{n} = \ou{m}{h}{n}\,\ou{0}{A}{0}\, \ou{m}{h}{n}{}^{-1}-d \ou{m}{h}{n}\, \ou{m}{h}{n}{}^{-1}=-d \ou{m}{h}{n}\, \ou{m}{h}{n}{}^{-1}\,.
\eeq
This connection is automatically flat since the curvature is $\ou{m}{F}{n}=\ou{m}{h}{n}\,\ou{0}{F}{0}\,\ou{m}{h}{n}{}^{-1} =0$. In general the tetrad extracted from the decomposition $\ou{m}{A}{n}=\ou{m}{w}{n}+\frac{1}{\ell}\ou{m}{E}{n}$ is very complicated. But the metric takes on a surprisingly simple form. To simplify the calculations, we first note that any $\ou{m}{h}{n}$ can be split into $\ou{m}{h}{n}=\ou{m}{h}{m} \ou{0}{h}{-q}$ where $q\equiv m-n$. We then notice that being diagonal $\ou{m}{h}{m} \in H =SU(2)_{diag}$. But the stabilizer subgroup is the subgroup that preserves the metric. Thus it is sufficient to calculate the metric for $\ou{0}{A}{-q}$ since gauge transforming the connection by $\ou{m}{h}{m}$ doesn't change the metric. The end result of this calculation is
\beq
\ou{m}{\bm{g}}{n}= \ell^2 \left(q^2\,d\chi^2 + \sin^2(q\chi) \left( d\theta^2 +\sin^2\theta \,d\phi^2 \right)\right)\,,
\eeq
and it only depends on $|q|=|m-n|$.

\subsection{Understanding the new geometries on the three sphere: geodesics and physical interpretation \label{NewGeometry}}
Our task is now to understand these new geometries on the three-sphere. From the perspective of the gauge theory these configurations were extremely natural (maybe unfamiliar, but not unprecedented...these tricks play an important role in most non-abelian gauge theories). What do they represent from the Riemmannian geometry perspective? Let me go into this in detail since it is bound to cause some confusion and controversy. I would like to first point out that we have only changed the geometry of the three-sphere, not its topology. What this means is that barring the usual, innocuous coordinate singularities at $\chi=0$ and $\chi=\pi$, the coordinates are good in the full range $0< \chi <\pi$, $0 <  \theta < \pi$, and $0\leq \phi < 2\pi$. I stress this because one might try to claim that the geometry is just the three-sphere geometry if we rescale $\chi\rightarrow \chi'=|q|\chi$. You can always do this, but then you must adjust the coordinate ranges to $0<\chi < |q| \pi$ to cover the whole three-sphere. So let's just stick with $\chi$.

With these coordinate ranges, we seem to have a problem. The metric appears to have pathologies at $\chi =\frac{a}{|q|}\pi$ for any integer $0<a<|q|$ (not including the coordinate singularities at the poles $\chi=0$ and $\chi=\pi$ which are just as innocuous as before). Looking at the determinant of the metric, or more generally the determinant of the triad (which will tell you something about the orientation of the triad) we have
\beq
det(\ou{m}{E}{n})=-\ell^3q \,\sin^2(q\chi) \,\sin(\theta)
\eeq
which goes to zero at these points. Of course, this could simply mean that we have just chosen the wrong coordinates, and a more judicious choice would remove these singularities. We need to compute a diffeomorphism invariant quantity if we really want to distinguish these states. Can we do this? We have the determinant of the triad, why not integrate it to get the volume of the three sphere? Here it is:
\beqa
{}^3\ou{m}{V}{n} &=& \int_{\mathbb{S}^3} \frac{1}{3!}\,\epsilon_{ijk} \,\ou{m}{E}{n}{}^i \w \ou{m}{E}{n}{}^j \w \ou{m}{E}{n}{}^k  \nn\\
&=& -\ell^3 q \int^\pi_{\chi=0}\int^\pi_{\theta=0}\int^{2\pi}_{\phi=0} \sin^2(q\chi) \,\sin(\theta)\,d\chi \w d\theta \w d\phi \nn\\
&=& -q\,2\pi^2 \ell^3 \,.
\eeqa
So the oriented volume of the new geometries on the three-sphere is just $-q$ times the volume of the ordinary three-sphere. But does this quantity really serve to distinguish the geometries as physically distinct? The key is that we need to show that the integral is invariant not only under $SU(2)_{diag} \rtimes Diff(\Sigma)$ (which you should already be convinced that it is) but invariant under the full group $Spin(4)\rtimes Diff(\Sigma)$. In fact it is -- let me calculate it in two different ways to convince you.

First, I want to show that the result of the integral is not dependent on some clever choice of coordinates but really does pick out some topological information that is invariant under diffeomorphisms. The integral can be related to the difference of two Chern-Simons integrals. Recall that $A^i_{\ua}=w^i+\frac{1}{\ell} E^i$ and $A^i_{\da}=w^i-\frac{1}{\ell} E^i$. The volume can then be related to the difference of the Chern-Simons functionals for $A_{\ua}$ and $A_{\da}$ by
\beqa
-{}^3\ou{m}{V}{n}\big/2\pi^2 \ell^3 &=& -\frac{1}{2\pi^2 \ell^3} \int_{\mathbb{S}^3} \frac{1}{3!}\,\epsilon_{ijk} \,\ou{m}{E}{n}{}^i \w \ou{m}{E}{n}{}^j \w \ou{m}{E}{n}{}^k \nn\\ 
&=& Y_{CS}[\ou{m}{\omega}{n}{}^i+\frac{1}{\ell}\ou{m}{E}{n}{}^i]-Y_{CS}[\ou{m}{\omega}{n}{}^i-\frac{1}{\ell}\ou{m}{E}{n}{}^i]\nn\\
&=& Y_{CS}[\ou{m}{A}{n}{}^i_{\ua}]-Y_{CS}[\ou{m}{A}{n}{}^i_{\da}] \nn\\
&=& \left( Y_{CS}[\ou{0}{A}{0}{}^i_{\ua}]+m \right)-\left(Y_{CS}[\ou{0}{A}{0}{}^i_{\da}]+n \right) \nn\\
&=& m-n \,.
\eeqa
Although the individual Chern-Simons functionals are not invariant under large gauge transformations, the {\it difference} of the two is invariant under the full group $Spin(4)\rtimes Diff(\Sigma)$. This also shows very explicitly how the volume is related to topological information about the flat connection.

Just one more integral to drive the point home and (hopefully) clear up any residual skepticism. An inquisitive reader may have been concerned with the following. As we said, all flat connections on $\mb{S}^3$ are gauge related. But all our connections are flat and I have constructed the geometries by gauge transforming the trivial $\ou{0}{A}{0}=0$ connection. So how is it that I could get new geometric configurations invariant under $Spin(4)\rtimes Diff(\Sigma)$ if they are all the $\ou{m}{A}{n}$ are gauge related? The answer is that the geometry is {\it not just} encoded in $\ou{m}{A}{n}$ but also in the choice of Cartan decomposition of $\ou{m}{A}{n}$ into $\ou{m}{w}{n}$ and $\ou{m}{E}{n}$. As explained previously, this choice is given by fixing a vector field $V^{\h{a}}$ living in $G/H=Spin(4)/SU(2)_{diag}$. Although we have not said it explicitly in the paragraphs above, the decomposition $A=w^i\,\tau_i +\frac{1}{\ell} E^i \,\eta_i$ actually does correspond to a choice of $V^{\h{a}}$, namely the choice $V^{\h{a}}=(0,0,0,1)$. Thus, what we really have been doing when we send $\ou{0}{A}{0} \rightarrow \ou{m}{A}{n}$ and extracting the triad is ``gauge transforming" the connection but {\it not} the vector field $V^{\h{a}}$. Such a transformation is {\it not a gauge transformation} at all, but a map to a new configuration. So the most explicit way to see that the volume is invariant under $Spin(4)\rtimes Diff(\Sigma)$ is to write the volume as a functional involving both fields $A$ and $V$. But I've already shown how to do this (in the context of $Spin(5)$, but the result generalizes). Here it is:
\beq
{}^3 V[A,V^{\h{a}}]\big/ 2\pi^2 \ell^3=\frac{1}{2\pi^2} \int_{\Sigma} \frac{1}{3!} \epsilon_{\h{a}\h{b}\h{c}\h{d}} \, V^{\h{a}} \,D_A V^{\h{b}} \w D_A V^{\h{c}} \w D_A V^{\h{d}}
\eeq
In the ``geometric gauge" $V^{\h{a}}=(0,0,0,1)$, this gives 
\beq
{}^3 V[A,V^{\h{a}}]\big/ 2\pi^2 \ell^3=\frac{1}{2\pi^2 \ell^3} \int_{\mathbb{S}^3} \frac{1}{3!}\,\epsilon_{ijk} \,\ou{m}{E}{n}{}^i \w \ou{m}{E}{n}{}^j \w \ou{m}{E}{n}{}^k 
\eeq
whereas in the ``topological gauge" ($A=0$) this gives
\beq
{}^3 V[A,V^{\h{a}}]\big/ 2\pi^2 \ell^3=\frac{1}{2\pi^2} \int_{\Sigma} \frac{1}{3!} \epsilon_{\h{a}\h{b}\h{c}\h{d}} \, V^{\h{a}} \,dV^{\h{b}} \w dV^{\h{c}} \w dV^{\h{d}}\,
\eeq
which is the winding number of $V^{\h{a}}:\mb{S}^3 \rightarrow \mb{S}^3$.
In any case for the flat connections the result of the integral is ${}^3 V[A,V^{\h{a}}]\big/ 2\pi^2 \ell^3=-q$, and it is invariant under $Spin(4)\rtimes Diff(\Sigma)$. So hopefully this puts any residual skepticism to rest: the geometry defined by two configurations of flat connections are physically distinct if $q_1 \neq q_2$.

Now let's return to the physical interpretation of the geometries. Within the ranges $\frac{a-1}{|q|}\pi< \chi < \frac{a}{|q|}\pi$ for integers $1\leq a \leq |q|$, the geometry just looks like the ordinary geometry on the three sphere. That is, observers used to observing the world on the backdrop of a metric would perceive that they are living in the ordinary three sphere geometry as long as they didn't get too close to the degenerate surfaces. From this geometric perspective, these surfaces would appear to be single points since they have zero volume. But this is just an artifact of the metric becoming degenerate at these points. An enlightened observer familiar with gravity or geometry as a secondary property of a more fundamental gauge theory would recognize that these surfaces are not single points at all but two-spheres where the pull-back of the metric is zero. So the geometry looks essentially like a string of pearls. It consists of a string of three-spheres attached to each neighboring three-sphere at a two-sphere where the metric becomes degenerate. From the topological point of view, the configuration might be described as follows. Given two $n$-dimensional manifolds $\Sigma_1$ and $\Sigma_2$, the $\#$-product or {\it connected sum} denoted $\Sigma_1 \# \Sigma_2$ is the space obtained by removing a ball $\mathbb{B}^{n}$ from each and gluing the two together along the $\partial \mathbb{B}^{n}=\mathbb{S}^{n-1}$ boundary. From the purely topological perspective, for any $n$-dimensional manifold $M$, we have $M\# \mathbb{S}^n \simeq M$. Thus, out geometries can be described by
\beq
\underbrace{\mathbb{S}^3\# \mathbb{S}^3 \# \cdots \# \mathbb{S}^3}_{|q|-times} \simeq \mathbb{S}^3\,.
\eeq
But we have the additional restriction that the metric determinant tends to zero, and the pull-back of the metric is identically zero on the two-spheres where the individual $\mathbb{S}^3$ spheres are glued. 

\begin{figure}
 \begin{center}
\includegraphics[height=6.0cm]{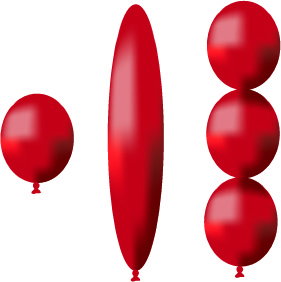}
   \end{center}
  \caption{\label{Baloons} A heuristic model for the exotic three sphere geometries. Start with a party balloon, and inflate it. Then over-inflate it. Twist the balloon into $|q|$-domains, each with surface area equal to the original.  Pictured above is the case $q=3$. This gives a heuristic picture of how an observer used to viewing physics from the metric perspective would view the geometry if they lived on the surface of the balloon.
  }
  \end{figure}

\begin{figure}
 \begin{center}
\includegraphics[height=6.0cm]{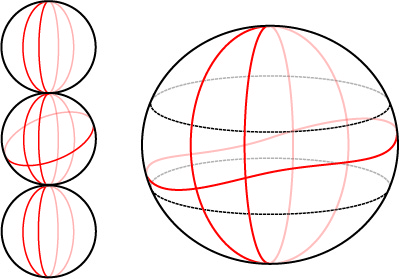}
   \end{center}
  \caption{\label{3Spheres} Here are two ways of picturing the exotic 3-spheres for the case q=3. The picture on the left represents what an observer used to observing physics from the metric perspective might see. Most notably, such an observer would view a two sphere with zero area as a single point. Thus, the degenerate two spheres are the single points joining the three 3-sphere domains. Pictured (in red) is one geodesic that is the ordinary great circle lying enirely within one 3-sphere domain, and two geodesics that travel along the great circles but through the degenerate surfaces into each of the 3-sphere domains. The picture on the right more accurately represents the actual geometry that an enlightened observer capable of viewing the interior of a surface with zero area would see. The dotted lines are the two degenerate 2-spheres where the metric determinant goes to zero. The red lines are the same geodesics as those on the left. Notice the two geodesics passing through the degenerate surfaces do not intersect except at the poles.}
\end{figure}

To clarify this picture let's look at the geodesics of the degenerate geometry. In fact, one can write the geodesic equation in a way that does not use the inverse of the metric so it is well defined even when the metric is degenerate. I could calculate the geodesics explicitly, but the result is easy to guess. Geodesics that do not cross the degenerate surfaces are precisely the ordinary geodesics on the three-sphere. Geodesics that live entirely within the degenerate two-spheres are completely unconstrained since the pull-back of the metric to these surfaces is zero. The most interesting geodesics are the ones that cross through the degenerate two surfaces. The only solutions that exist are solutions where the tangent to the curve is orthogonal to the two-surface at the point where the geodesic crosses the surface. These curves extend into the neighboring three-sphere regime and can be interpreted as the great circles of the sphere, as shown in Fig. (\ref{3Spheres}). 

So from the perspective of an observer used to dealing with the metric description of geometry, one or both of the poles of the three-sphere will look like a point that acts as a portal into the neighboring regime, and allows for communication between the neighboring three-spheres.

\subsection{Weird stuff happens when you change the gauge}
Before moving on to the de Sitter theory, let me make one more detour to illustrate some of the extreme consequences of the full gauge group being a combination of $Spin(4)$ and diffeomorphisms. In the previous section I choose a clever gauge where although the tetrad may look somewhat complicated, the metric and its degeneracy structure was as simple as it could possibly be. This may have given the impression that the associated geometries are somewhat trivial. But by gauging and diffeomorphing the manifold, one can push around the degenerate surface and contort them, or even change their character altogether. To illustrate this, let me choose another gauge where the richness of the geometry may become more apparent. This will serve to demonstrate yet again how different things can look in different gauge, and why it is was so important to construct a topological invariant to distinguish the different states. 

\begin{figure}
 \begin{center}
\includegraphics[height=15cm]{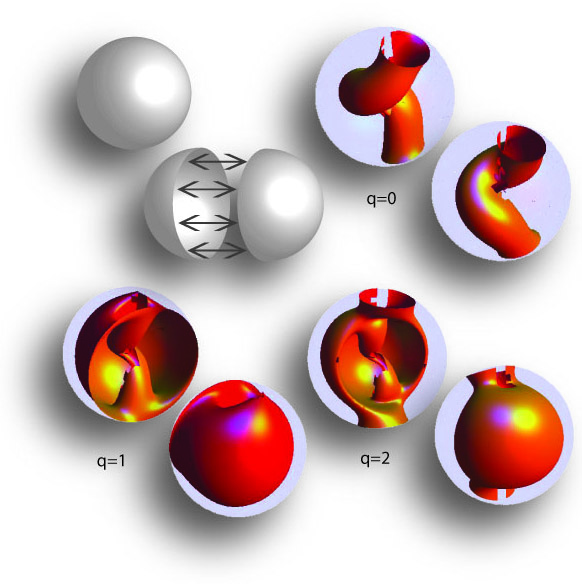}
   \end{center}
  \caption{\label{Degenerates} Pictured here are the degenerate surfaces for $q=0$, $q=1$, and $q=2$. The three sphere is visualized by cutting it along a two-sphere boundary to form two closed balls, and identifying the boundaries of the two. The irregularities in the plotted surfaces are caused by numerical sampling errors. When the boundaries are identified, it is clear that the resulting degenerate 2-surface has genus $|q+1|$. Images were created using Mathematica\textregistered.}
\end{figure}

Suppose that given the ordinary three-sphere geometry, we had {\it chosen} the triad to be the diagonal triad
\beq
E^1=\ell \,d\chi \quad\quad E^2=\ell\, \sin(\chi) d\theta \quad \quad E^3=\ell \,\sin(\chi) \sin(\theta) \,d\phi\,.
\eeq
Call this triad and its corresponding torsion-free spin connection $(\ou{0}{E}{1}{}_*,\ou{0}{w}{1}{}_*)$ and the associated $Spin(4)$ connection $\ou{0}{A}{1}{}_*=\ou{0}{w}{1}{}_*+\frac{1}{\ell}\ou{0}{E}{1}{}_*$. Now, let's start with this connection instead of the zero connection, and just as before build in infinite tower of flat connection $\ou{m}{A}{n}{}_*$. Then extract the tetrad and the induced metric $\ou{m}{\bm{g}}{n}{}_*$. This has been done in \cite{RandonodSSpaces}. The resultant metric, not surprisingly still has degeneracies. But, the degenerate surfaces look very different. I plot them in Fig. (\ref{dSSpaces}). As opposed to there being $|q|$ distinct degenerate surfaces each with topology $\mathbb{S}^2$, there appears to be only one degenerate surface regardless of $|q|$ (except for the original $\ou{0}{\bm{g}}{1}{}_*$ which has no degenerate surfaces). Moreover, the topological information about the quantum number $q$ appears to be encoded not only in the volume integral, which still gives the same answer since it is a topological quantity invariant under $Spin(4)\rtimes Diff(M)$, but also in the {\it genus} of the two-surface as seen in Fig. (\ref{Degenerates}). Although I have not been able to prove anything rigorous, it appears, and I feel confident conjecturing, that the genus of the two-surface has genus given by\footnote{Watch for the change of conventions for $q$ from \cite{RandonodSSpaces}. There I began with the ordinary geometry and constructed the connection so it was natural to call it $\ou{0}{A}{0}$. The relation between the two is $m_{old}=m_{new}$ and $n_{old}=n_{new}-1$ so $q_{old}=q_{new}+1$.}
\beq
\text{Genus(Degenerate Surface)}\stackrel{?}{=}|q+1|. \label{Conjecture}
\eeq

This is an extreme example of the enormous degree of freedom contained in $Spin(4)\rtimes Diff(M)$. The richness of the diffeomorphism group should be familiar from general relativity, but the internal gauge group adds one more layer of complexity. As pointed out in \cite{Mielke:1992gk}, the internal translations can be used to add or remove apparent metric degeneracies in isolated regions. However, in our example it appears likely that there are topological obstructions to removing the metric degeneracies entirely. But, it is clear from this example that the character of the degenerate surfaces will be very different in different gauges.

\section{Extension to de Sitter \label{dSExtension}}
Now I want to return to the de Sitter gauge theory. The reason why we took the detour into the geometries on the three-sphere is that most of the major results generalize to the de Sitter case. There are two reasons for this, which I have already discussed but will repeat here. First there is the topology of de Sitter space itself, $dS_4\simeq \mb{R}\times \mb{S}^3$. A typical spatial slice has the topology and geometry of the constant curvature three sphere. Next there is the structure of the gauge group $Spin(4,1)=\overline{SO}(4,1)$. The maximal compact subgroup form the de Sitter group is $Spin(4)=SU(2)\times SU(2)$, and the remaining part from the topological perspective is not as interesting. Thus, many of the results of the previous section generalize.

First I want to introduce some notation and conventions for the $Spin(4)$ gauge theory. It will be easiest to work with the Clifford algebra notation. As in the previous section, I will assume that the involution vector field $V^{A}$ is trivial so that $V^A=(0,0,0,0,1)$. This is the easiest way to compare different geometries extracted from the gauge theory. Recall that in this gauge, the connection can be split as 
\beq
\cA=\omega+\frac{1}{\ell} \g_5 e=\omega^{IJ}\,{\ts \frac{1}{4}}\g_{[I}\g_{J]} + \frac{1}{\ell} e^I \,{\ts \frac{1}{2}}\g_5\g_I
\eeq

Let's work in the Dirac representation for the Clifford algebra given by (recall we are using $(-,+,+,+)$ signature so there are some extra $i$'s floating around)
\beqa
\gamma^0=-i\left[\begin{array}{cc} 1 & 0 \\ 0 & -1 \end{array}\right] \quad \quad 
\gamma^i=-i\left[\begin{array}{cc} 0 & \sigma^i \\ -\sigma^i & 0 \end{array}\right] \quad \quad
\gamma_5=\left[\begin{array}{cc} 0 & 1 \\ 1 & 0 \end{array}\right] \label{DiracRep}
\eeqa
so that
\beqa
\frac{1}{2}\gamma^{[i}\gamma^{j]}=\left[\begin{array}{cc} \frac{i}{2}\epsilon^{ij}{}_k\,\sigma^k & 0 \\ 0 & \frac{i}{2}\epsilon^{ij}{}_k\,\sigma^k \end{array}\right] \quad \quad
\frac{1}{2}\gamma_5 \gamma^k=\left[\begin{array}{cc} \frac{i}{2}\sigma^k & 0 \\ 0 & -\frac{i}{2}\sigma^k \end{array}\right]\ .
\eeqa
Now define
\beqa
\tau^i_\ua &\equiv&  \frac{1}{2}\left (\frac{1}{4}\epsilon^i{}_{jk}\gamma^{[j}\gamma^{k]}+\frac{1}{2} \gamma_5 \gamma^i \right)=\left[\begin{array}{cc} \frac{i}{2}\sigma^i & 0 \\ 0 & 0 \end{array}\right] \\
\tau^i_\da &\equiv&  \frac{1}{2}\left(\frac{1}{4}\epsilon^i{}_{jk}\gamma^{[j}\gamma^{k]}-\frac{1}{2} \gamma_5 \gamma^i \right) =\left[\begin{array}{cc} 0 & 0 \\ 0 & \frac{i}{2}\sigma^i \end{array}\right]\ .
\eeqa
These are of course the same generators that we defined in the previous section, so they form the generators of the maximal compact subgroup $Spin(4)=SU(2)_{\ua}\times SU(2)_{\da}$. On thing to notice is that they are composed of linear combinations of a spatial rotation (in the stabilizer algebra $\mf{h}=\mf{su}(2)$) and a spatial translation (not in the stabilizer algebra). This is the reason why this stuff works for de Sitter and does not work for anti-de Sitter or Poincar\'{e}.

Now we can construct group elements in the large sector of $Spin(4)\subset Spin(4,1)$ in exactly the same way as before:
\beq
\ou{1}{h}{0} =\left[\begin{matrix} Y_{\h{4}}\bm{1} + Y_{\h{i}} \,i\sigma^i & 0 \\ 0 & \bm{1} \end{matrix} \right] \quad \quad \ou{0}{h}{1} =\left[\begin{matrix} \bm{1} & 0 \\ 0 & Y_{\h{4}}\bm{1} + Y_{\h{i}} \,i\sigma^i  \end{matrix} \right]\,.
\eeq
and 
\beq
\ou{m}{h}{n}\equiv \ou{1}{h}{0}{}^m\,\ou{0}{h}{1}{}^n\,.
\eeq

The strategy is to build the connection defining de Sitter space in a two step process. Suppose we had a triad on the three sphere $E^i$ which we extended to the four dimensional manifold by setting $E^i_t=0$. Then a tetrad on $M=\mb{R}\times \mb{S}^3$ could be defined by
\beq
e^0= dt \quad \quad e^i=\cosh(t/\ell)\, E^i \,.\label{Tetrad1}
\eeq
Assuming $E^i$ defines the constant curvature, zero torsion geometry on the 3-sphere, then the induced metric would be
\beq
\bm{g}=\eta_{IJ} \,e^I \otimes e^J =-dt^2 +\ell^2 \cosh^2(t/\ell) \left( d\chi^2 + \sin^2\chi\left(d\theta^2+\sin^2\theta\, d\phi^2\right) \right)
\eeq
which is of course the de Sitter metric. So the strategy is to use what we already know to construct the triad from a gauge transformation of the trivial connection, and then time-translate the connection using another group element in $Spin(4,1)$ to try to reproduce (\ref{Tetrad1}).

In fact this is not that difficult. First define the connection $\ou{0}{A}{1}=-d\ou{0}{h}{1}\,\ou{0}{h}{1}{}^{-1}$. This does not yet define a proper geometry on $M$ so we have called it $\ou{0}{A}{1}$ as opposed to $\ou{0}{\cA}{1}$. Now the second step is to time translate this connection. Take our time translation to be $h_t \equiv \exp(\frac{1}{2} t \,\gamma_5 \gamma^0)$. Note that this is a {\it small} gauge transformation since it is deformable to the identity so it will not affect the winding numbers of the group. The connection then becomes
\beq
\ou{0}{\cA}{1}=h_t\,\ou{0}{A}{1}\,h^{-1}_t-dh_t\,h^{-1}_t=-(d(h_t \ou{0}{h}{1}))(h_t \ou{0}{h}{1})^{-1}\,.
\eeq

It takes a little algebra, but it can be checked that when the tetrad is extracted from the Cartan decomposition $\ou{0}{\cA}{1}=\ou{0}{\omega}{1}+\frac{1}{\ell} \g_5 \,\ou{0}{e}{1}$, it turns out to be exactly (\ref{Tetrad1}) with $E^i=\ou{0}{E}{1}{}^i$ given by the triad on the 3-sphere defined in the previous section.

We can now extend this procedure to construct more exotic geometries on $M$, just as we did on the three-sphere. All you have to do is first define $\ou{m}{A}{n}=-d\ou{m}{h}{n}\,\ou{m}{h}{n}{}^{-1}$ and then time translate by $h_t$. But, to make things a bit cleaner, it is would be nice if we could just take powers of some group element instead of going through this two-step process. So I will define $\ou{0}{g}{1}\equiv h_t\, \ou{0}{h}{1} \,h_t^{-1}$ and $\ou{1}{g}{0}\equiv h_t \,\ou{1}{h}{0} \,h_t^{-1}$, and of course $\ou{m}{g}{n}=\ou{1}{g}{0}{}^{m}\,\ou{0}{g}{1}{}^n$. Instead of starting with the zero connection start with 
\beq
\ou{0}{\cA}{0}\equiv -dh_t \, h_t^{-1}\,.
\eeq
The tetrad extracted from this is $e^0=dt$ and $e^i=0$. The full set of connections then is given by
\beq
\ou{m}{\cA}{n}=\ou{m}{g}{n} \,\ou{0}{\cA}{0} \,\ou{m}{g}{n}{}^{-1}-d\ou{m}{g}{n}\,\ou{m}{g}{n}{}^{-1}=-(d(h_t \ou{m}{h}{n}))(h_t \ou{m}{h}{n})^{-1}\,.
\eeq
It should be clear that all these connections are flat ($\ou{m}{F}{n}=0$) since they are ``gauge" related to the zero connection. As before, the extracted tetrad $\ou{m}{e}{n}{}^I$ is very complicated. But the metric induced from it is very simple. You can probably guess what the result is:
\beqa
\ou{m}{\bm{g}}{n} &\equiv& \eta_{IJ} \,\ou{m}{e}{n}{}^I \otimes \ou{m}{e}{n}{}^J \nn\\
&=& -dt^2+\ell^2\cosh^2(t/\ell)\,\left(q^2\,d\chi^2 +\sin^2q\chi\,\left(d\theta^2 +\sin^2{\theta} \,d\phi^2\right) \right)\,, \label{GeneralizeddS}
\eeqa
where again $q\equiv m-n$.

\subsection{Interpretation of the solutions \label{Interpretation}}
Now let's try to understand what these geometries mean. As before, the solutions can be distinguished by their topological number $q$, which is directly related to the spatial volume. The procedure for constructing a topological invariant for these configurations is mostly identical to that of the three-sphere \cite{RandonodSSpaces}. The major caveat is that one has to identify a preferred spatial slice. This can be accomplished by noting that all of the solutions have a time-reversal symmetry, and the distinguished spatial slice $\Sigma_0$ can be identified with the spatial slice that is left invariant under time reversal. The pull-back of the connection to the spatial slice $\Sigma_0$ is just the connection $\ou{m}{A}{n}$ on the three-sphere defined in the previous section, and all the arguments still hold. Thus, the topological number is 
\beq
q=-{}^3\ou{m}{V}{n}(\Sigma_0) \big/2\pi^2 \ell^3
\eeq

\begin{figure}
 \begin{center}
\includegraphics[height=3.5cm]{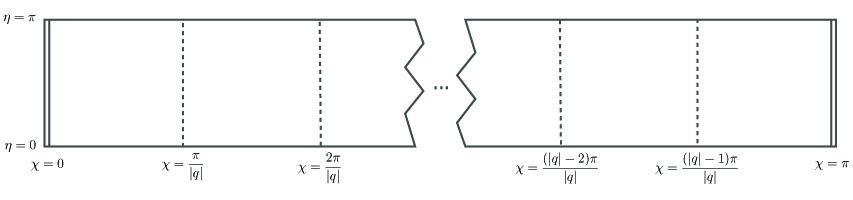}
   \end{center}
  \caption{\label{Conformal} Pictured here is the conformal diagram for the generalized de Sitter spacetime for arbitrary $q$. It consists of a set of de Sitter blocks glued together along two-spheres (the dotted lines) where the metric determinant goes to zero. The left and right edges of the diagram are the poles at $\chi=0$ and $\chi=\pi$ or $\chi'=0$ and $\chi'=|q|\pi$.}\end{figure}

As with the three-dimsensional case, the geodesics can be found explicitly through the geodesic equation, but the result can be guessed from the previous analysis. Just as before, within a three sphere domain $\frac{a-1}{|q|}\pi < \chi < \frac{a}{|q|}\pi $, the geodesics that do not pass through the poles are identical to those of ordinary de Sitter space. The main difference is the replacement of the poles of the three-sphere domains (other than the points $\chi=0$ and $\chi=\pi$) with degenerate two spheres connecting one domain to the next. Timelike, spacelike, and lightlike geodesics can pass through these points. Just as before, the only geodesics passing through the degenerate two-sphere to emerge into another three-sphere domain must be ``perpendicular" to the sphere, i.e. they must have $\dot{\chi}\neq 0$ and $\dot{\theta}=\dot{\phi}=0$. Rather than consider explicit geodesics, it is more useful to consider the conformal diagram of the resulting metric. To get this, make the coordinate transformation $\{t,\chi,\theta,\phi\}\rightarrow\{t',\chi',\theta,\phi\}$ where
\beq
\cosh(t/\ell)=\frac{1}{\cos(t')} \quad\quad \quad \chi'=q\chi\,.
\eeq
In these coordinates the metric takes the form
\beq
\ou{m}{\bm{g}}{n}=\frac{\ell^2}{\cos^2t'}\left(-dt'^2 +d\chi'^2+\sin^2\chi'\left(d\theta^2+\sin^2\theta \,d\phi^2\right)\right)
\eeq
which is conformal to the Einstein static universe. The coordinates now take the ranges $-\pi/2 <t' <\pi/2$ and $0\leq \chi' \leq q\pi$ subject to the ordinary coordinate singularities at $\chi'=0$ and $\chi'=q\pi$. The conformal diagram is shown in Fig. (\ref{Conformal}). It very clearly shows how the geometries can be interpreted as $|q|$-copies of de Sitter space glued together at their poles into strings of three spheres that behave much like ordinary de Sitter space. The major exception being that one copy can communicate with its nearest neighbor through the poles. 

\subsection{Common questions and misconceptions}
Let me now take some time to address so common questions and misconceptions I have encountered regarding the construction I have outlined above. Some of these points have been addressed already, but no harm is done in repeating them.
\begin{itemize}
\item{{\it \textbf{All the geometries you have constructed are gauge related, so in a true de Sitter gauge theory where the full group is retained, aren't all the configurations on the same gauge orbit, and therefore just different ways of writing the same thing?:}} No. There are two ways of viewing this, the explicit symmetry breaking scenario and the (quasi)-dynamic symmetry breaking scenario. In the explicit symmetry breaking scenario we simply break the symmetry by hand by introducing some fixed field that breaks $Spin(4,1)\rightarrow Spin(3,1)$. In this case all the configurations $\ou{m}{\mathcal{A}}{n}=\ou{m}{\omega}{n}+\frac{1}{\ell}\ou{m}{e}{n}$ are on a $Spin(4,1)$ gauge orbit, but the gauge group of the theory is not $Spin(4,1)$, it is $Spin(3,1)$. Analyzed with respect to this group, the solutions are not on a $Spin(3,1)$ gauge orbit. Thus, they are physically distinct. 

In the dynamic symmetry breaking mechanism the order parameter $V$ in $G/H$ breaks the symmetry spontaneously. The full gauge symmetry is retained, but after the symmetry breaking, it is retained in a non-obvious, generally non-linear way. In this case, the full configuration is {\it not} characterized by $\mathcal{A}$ alone, but by the pair $\{\mathcal{A},V\}$. A gauge transformation would take $\{\mathcal{A},V\}\rightarrow \{g\mathcal{A}g^{-1}-dg \,g^{-1},gVg^{-1}\}$. But as we have constructed the solutions, the procedure is to start with $V=V_*=(0,0,0,0,1)$ and $\mathcal{A}=\ou{0}{\mathcal{A}}{0}$, and construct the new configurations by the map $\{\ou{0}{\mathcal{A}}{0}, V_*\} \rightarrow \{\ou{m}{\mathcal{A}}{n}, V_*\}=\{\ou{m}{g}{n} \ou{0}{\mathcal{A}}{0} \ou{m}{g}{n}{}^{-1} -d\ou{m}{g}{n} \, \ou{m}{g}{n}{}^{-1}, V_*\}$. Thus one of the fields is ``gauge"-transformed while the other is not. The result is {\it not} a gauge transformation. So, the configurations are physically distinct even with respect to the larger gauge group $Spin(4,1)$.}

\item{{\it \textbf{The full gauge group is not $\mathbf{Spin(4,1)}$ or $\mathbf{Spin(3,1)}$ but it is $\mathbf{Spin(4,1)\rtimes Diff_M}$ or $\mathbf{Spin(3,1)\rtimes Diff_M}$. How do you know that these are not the same solutions just written in different coordinates?:}} Agreed, this is a very important issue. This is why it was so important to come up with a topological invariant, invariant under the full gauge group with diffeomorphisms, to pick out the charge. As we saw in the previous section, the combination of gauge transformations and diffeomorphisms can do radical things to your geometries, even completely changing the character of the degenerate surfaces. But, the charge we have constructed is invariant under the full gauge group including diffeomorphisms, so we know they are physically distinct.}

\item{{\it \textbf{The configurations you have given are such that the Ricci scalar blows up as you approach the degenerate surface. Thus, these points should be treated like any other singular points where GR breaks down, and they should be excised appropriately:}} Not exactly. The only problem with our configurations is that the metric determinant goes to zero at certain points. Thus, inverse metrics no longer exist everywhere. But, any expression that you can write down without using the inverse of the metric is not only finite, but continuous and differentiable. For example, it is true that the Ricci scalar proper blows up. But that is only because in order to write down the Ricci scalar you have to invert the metric. A more appropriate expression is $\epsilon_{IJKL}e^I \w e^J \w R^{KL}$ which is proportional to $Ricci \sqrt{|g|} d^4 x$ wherever the metric is invertible. The former is explicitly finite but doesn't require the inverse metric to make sense. As a consequence whereas the Ricci scalar blows up as we approach the degenerate surfaces, the combination $Ricci \sqrt{|g|}$ is always finite. 

One more point -- from the perspective of the quantum theory, it is finiteness of the action that really matters. If the action is finite on a configuration, the configuration is in some sense good. If it blows up, then we have a problem. But the action does not require the existence of an inverse metric or tetrad to write down, and it is explicitly finite on these configurations (in fact identically zero). So, we are safe.} 

\item{{\it \textbf{The degenerate surfaces are points where the volume goes to zero. So aren't these either coordinate singularities or just single points?:}} No. As we have emphasized, taking the gauge field to be primary and the metric to be secondary allows for solutions where the metric is degenerate over extended points or submanifolds. This allows for the possibility that extended chunks of spacetime can exist with zero volume. In our case, the metric (\ref{GeneralizeddS}) has degenerate surfaces with topology $\mathbb{R}\times\mathbb{S}^2$ which are extended submanifolds embedded in $M$ with zero volume. It is true that the location and even the topology of the individual degenerate surfaces can be changed by a combination of diffeomorphisms and gauge transformations. But, it is unlikely that the surfaces can be removed entirely as there is strong evidence that existence of degenerate surfaces is topologically ensured, as hinted at in our conjecture (\ref{Conjecture}).}

\item{{\it \textbf{Aren't these geometries the three-sphere with different points identified, or orbifolds, or algebraic varieties...?}} I have heard all sorts of attempts at interpreting the geometries as orbifolds, Lenz spaces, algebraic varieties, etc... These are usually based on preconceptions about metric geometries. Let me emphasize here that the gauge field is primary whereas the metric is secondary. Suppose I had given you a gauge field $B$ based on some group and I told you that you could combine some components of the field to produce a tensor $b_{\mu\nu}$ that happens to be symmetric in its two indices. And furthermore, the connection $B$ is almost as simple as it can get since it is gauge related to the flat connection. Then nobody would complain. I'm just putting a simple, smooth, continuous, differentiable, regular connection on $M$, just like we always do. Then suppose I relabelled the connection $B$, called it $\mathcal{A}$, split it such that $A=\omega+\frac{1}{\ell} e$, defined $b_{\mu\nu}=\eta_{IJ} e^I_\mu e^J_\nu$, and relabelled it $g_{\mu\nu}$. Nothing has changed. The field configuration is still smooth, continuous, differentiable, and regular. The topology has not changed. If I were then to call $g_{\mu\nu}$ a metric, the only problem would be that it is non-invertible at certain points. But that is fine, the metric is secondary, the gauge field is primary.} 
\end{itemize}

%%%%%%%%%%%%%%%%%%%%%%%%%%%%%
%%%%%%%%%%%%%%%%%%%%%%%%%%%%%

\section{Quantum gravity from the de Sitter gauge theory \label{QuantumGravity}}
We are now in a position to see the real advantage of the gauge approach to gravity. Up to this point, a critical reader may have objected that we put in an awful lot of work for a very simple result. After all, the solutions $\ou{m}{\bm{g}}{n}$ could have easily been guessed without ever making reference to the de Sitter gauge theory, and no one would have been very impressed since they do not appear to be physically relevant. But that's exactly the point! Solutions which in the ordinary metric formulation may be tossed aside as unphysical are perfectly natural within the context of gravity as a gauge theory. Moreover, they are not unprecedented -- they are the precise analogs of the degenerate Yang-Mills vacua for example. As in QCD, the real advantage of exploring these sectors comes from the quantum theory. Thus, we now turn to quantum gravity. Readers who have little knowledge of quantum gravity itself need not worry, since little knowledge of quantum gravity is necessary to understand these arguments.

\subsection{The multitude of vacua \label{Vacua}}
We have seen that the de Sitter gauge theory allows for an infinite class of solutions each with a high degree of symmetry corresponding to a flat connection. Such highly symmetric flat solutions are candidate solutions for the quantum ``vacuum" or ``ground state". I added scare quotes since I will later argue that the corresponding states are not actually stable ground states. Typically such vacuum-like candidates have a precise analogue in the quantum theory, often emerging as a lowest order WKB state that happens to also be an exact solutions to the quantum theory. In this sense, they are quantum states that are about as classical as a quantum state can be in the sense that there is a clear correspondence between the quantum state and a corresponding classical field configuration. The quantum state is represented by a vector in the Hilbert space $|\psi\rangle$. Let's assume that the Hilbert space of the quantum theory that we build from the de Sitter gauge theory retains a faithful representation of the local de Sitter gauge group. 

Now, starting with the flat configuration $\{\ou{0}{\cA}{0},V_* \}$ where we have used our gauge freedom to transform $V^{\h{a}}$ to $V^{\h{a}}_*=(0,0,0,0,1)$, we can build the tower of flat connections by a map $\ou{m}{G}{n}$ taking $\{\ou{0}{\cA}{0},V_* \}$ to $\ou{m}{G}{n}\left( \{\ou{0}{\cA}{0},V_* \} \right) =\{\ou{m}{\cA}{n},V_* \}$. Note, this is obvious but subtle -- the map is {\it not} a gauge transformation of the total field configuration (since it transforms $\cA$ but not $V$), though it is a faithful representation of the group of large gauge transformations, namely $\pi_3(Spin(4,1))=\mb{Z}\oplus \mb{Z}$. We should expect that such a map also exists as a quantum operator on the Hilbert space -- call it $\ou{m}{\mathcal{G}}{n}$. Using this map, we can build the tower of states dividing the Hilbert space, just like the configuration space, into sectors. Starting with the quantum state $|0,0\rangle$, which is the quantum state corresponding to the configuration $\{\ou{0}{\cA}{0},V_*\}$, one can build an infinite tower of states $|m,n\rangle=\ou{m}{\cG}{n}|0,0\rangle $ corresponding to the classical configurations $\{\ou{m}{\cA}{n}, V_*\}$. The de Sitter ground state can be identified with the state $|0,1\rangle$. 

\subsection{Instantons \label{Instantons}}
Possibly the most exciting new feature that could come about from quantum gravity based on the de Sitter gauge over and above the more standard quantization of Einstein-Cartan gravity is the possibility of quantum mechanical tunneling between different vacua. Take two candidate ground states $|m_i,n_i\rangle$ and  $|m_f,n_f\rangle$. Suppose we want compute the quantum mechanical transition amplitude from the initial state $|m_i,n_i\rangle$, which we might as well take to be the ground state in the infinite past, to the final state $|m_f,n_f\rangle$. In the canonical approach to quantum gravity, such a transition amplitude would be represented by the inner product $\lim_{t\rightarrow \infty}\langle m_f,n_f| \mathcal{T}\exp (it\hat{\mathcal{H}}) |m_i,n_i \rangle $. Due to subtleties regarding the problem of time in quantum gravity, this problem is generally transmuted into finding solutions to the quantum Hamiltonian constraint and an appropriate inner product. In the path integral picture, the transition amplitude can be computed by a sum over field configurations that are fixed at asymptotic past and future to be $\ou{m_i}{\cA}{n_i}$ and $\ou{m_f}{\cA}{n_f}$. In the path integral approach the transition amplitude is represented by 
\beq
\langle m_f ,n_f,\infty  | m_i,n_i,-\infty \rangle = \int\limits^{\mathcal{P}(\{\ou{m_f}{\cA}{n_f},V_*\})}_{\mathcal{P}(\{\ou{m_i}{\cA}{n_i},V_*\})} \mathcal{D}\cA \mathcal{D}V \, e^{i S[\cA, V]}\,,
\eeq
where the integration measure is assumed to be invariant under $Spin(4,1)\rtimes Diff(M)$, and $\mathcal{P}(\{\mathcal{A},V\})$ is a polarization of the phase space as dictated by the canonical theory\footnote{This is more familiar than it sounds. For example in ordinary quantum mechanics the transition $\langle q_f| q_i\rangle$ is computed in the path integral by integrating over all paths that start at $q_i$ and end at $q_f$ but puts no restriction on $p_i=\dot{q}_i$ and $p_f=\dot{q}_f$. This is a choice of polarization of the phase space $\{q,p\}$ to obtain a configuration space given by $\mathcal{P}(\{q,p\})=q$.}.

Why should we expect that the transition amplitude is non-zero? There are two reasons for this. First, there are analogous constructions in Yang-Mills theory or QCD. Take $SU(2)$ Yang-Mills theory as an example. Since $\pi_3(SU(2))=\mb{Z}$, the set of flat configurations also splits into sectors labelled by, in this case, a single integer. The Hilbert space also decomposes into a set of candidate ground states $|n\rangle$. In this case, there is an elaborate set of tricks \cite{Rajaraman:Instantons} (unfortunately most of which do not carry over to the de Sitter gauge theory) that allow for an approximation of the transition amplitude $\langle n_f|n_i \rangle$. This process should be thought of as a quantum mechanical tunneling between two inequivalent candidate ground states. The transition is known as an instanton\footnote{In some circles, what was once a trick, the Wick transform to the Euclidean theory to compute classical configurations that dominate the path integral, became a definition. So often one refers to an instanton as a classical solution to the Euclidean theory connecting different n-sectors. Here I will stick with the moniker, instanton, as referring to the quantum mechanical tunneling amplitude in the Lorentzian theory.}. The main trick is to perform a Wick rotation to imaginary time so that the manifold becomes four-dimensional Euclidean space with a Euclidean metric. In the Euclidean theory, the field equations admit exact {\it classical} configurations that connect the two degenerate vacua (roughly speaking, this is because the potential flips sign in the transition to the Euclidean theory). These will dominate the Euclidean path integral allowing for approximate calculations of the Euclidean transition amplitude, which do turn out to be non-zero.

It would be nice if we could use these tricks in the de Sitter case, but unfortunately it doesn't look like it is possible (or at least it is not obvious how). The reason is because in the gravitational case, first of all since the spacetime is curved, the Wick transformation is not well defined or understood properly. Secondly, the signature of the metric is intimately tied with the structure of the gauge group. In order to Euclideanize the theory, you also have to Euclideanize the group. This means the group $G=Spin(4,1)$ with stabilizer subgroup $H=Spin(3,1)$ becomes $G=Spin(5)$ with stabilizer group $Spin(4)$. But, $Spin(5)$ and $Spin(4)$ have very different topological properties from $Spin(4,1)$ and $Spin(3,1)$, and the analogous geometric structures do not exist in the Euclidean case.

\begin{figure}
 \begin{center}
\includegraphics[height=11cm]{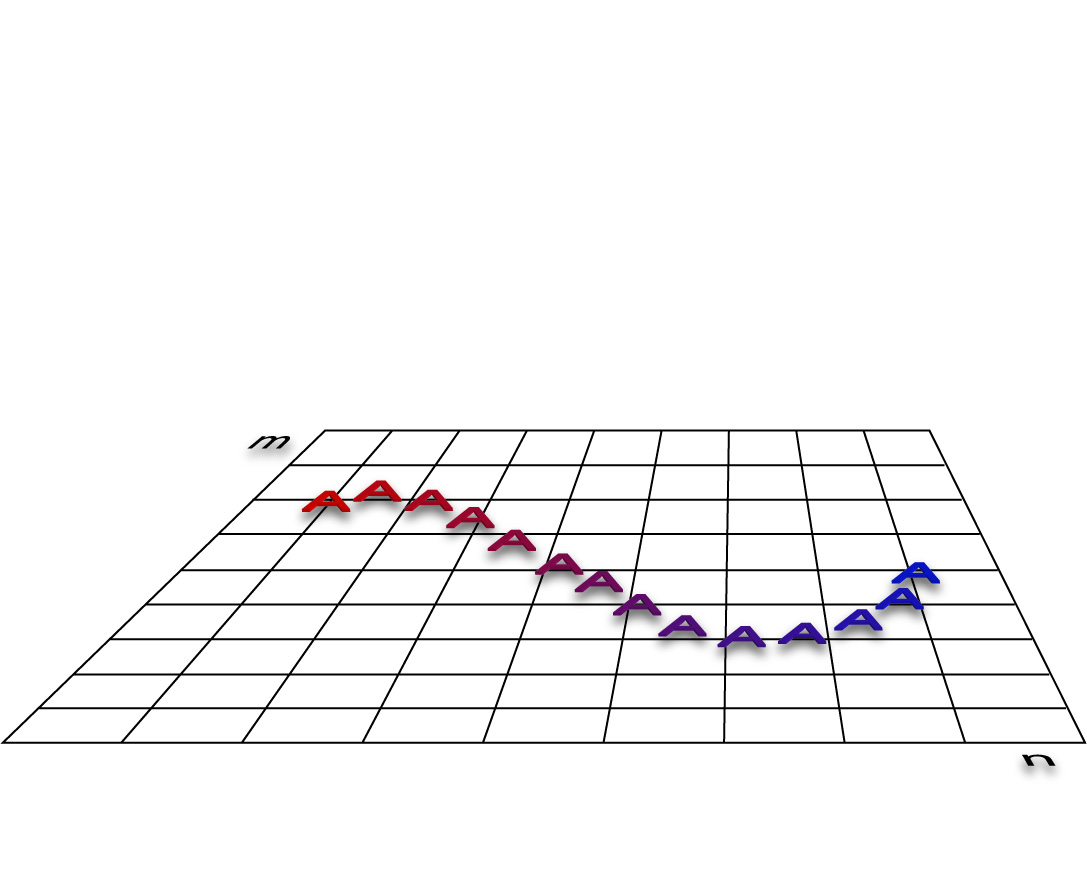}
   \end{center}
  \caption{\label{As} This is a schematic picture illustrating the path connectedness of the space of connections. Each vertex on the graph represents a topologically distinct flat connection. Starting with a smooth connection in an $(m_i, n_i)$ the smooth path interpolates between this and the end point in the $(m_f, n_f)$ sector. The path necessarily passes over regions where the connection is not flat.}
\end{figure}

On the other hand, the key property that makes the {\it Lorentzian} transition amplitude, viewed as a sum over paths, non-zero in the Yang-Mills case {\it does} carry over to the de Sitter case. Actually its just a generic property of theories based on a space of connections, and this is one of the primary advantages to viewing gravity as a gauge theory. The key property is the existence of smooth, continuous paths that connect the two degenerate vacua. There are no smooth paths that connect $\{\ou{m_i}{\cA}{n_i},V_*\}$ to $\{\ou{m_f}{\cA}{n_f},V_*\}$ (for $\{m_f,n_f\}\neq \{m_i,n_i\}$) that live entirely within the space of flat connections. This is impossible from topological considerations since all flat connections are gauge equivalent but $\ou{m_i}{g}{n_i}$ is homotopically distinct and can never be deformed into $\ou{m_f}{g}{n_f}$ by a continuous transformation. But, there {\it are} continuous and smooth paths that are not flat but still connect the two solutions (see Fig. (\ref{As})). 

This is because the space of connections forms an affine space which allows any point (configuration of $\cA$) to be connected to any other point by a simple translation (in this case the addition of a tensor). More specifically, given any connection $\cA_1$, I can always connect it to $\cA_2$ by addition of tensor $C$, so that $\cA_2=\cA_1+C$. Of course the tensor is just $C=\cA_2-\cA_1$, which transforms homogenously under a gauge transformation since
\beq
{}^gC=(g\cA_2 g^{-1} -dg g^{-1})-(g\cA_1 g^{-1} -dg g^{-1})=gCg^{-1}.
\eeq
A parameterized smooth path connecting the two configurations can easily be constructed. Let $s$ range smoothly and monotonically from $0$ to $1$ and define $\cA(s)=\cA_1+s(\cA_2-\cA_1)$. This is a smooth path in the space of connections starting at $\cA(s=0)=\cA_1$ and ending at $\cA(s=1)=\cA_2$. So suppose we wanted to construct a smooth path from $\ou{m_i}{\cA}{n_i}$ at $t=-\infty$ to $\ou{m_f}{\cA}{n_f}$ at $t=\infty$. There are lot's of ways to do this. Here's one: take $f(t)$ to be any smooth function monotonically increasing from $f(-\infty)=0$ to $f(\infty)=1$. Let's also assume $\frac{\p f}{\p t}=0$ at $t=\pm \infty$. Then define the connection
\beq
\cA_*(t)\equiv \ou{m_i}{\cA}{n_i} +f(t)(\ou{m_f}{\cA}{n_f}-\ou{m_i}{\cA}{n_i})\,.
\eeq
Clearly this is a smooth path in the space of connections parametrized by time, connecting the initial configuration $\cA_*(-\infty)= \ou{m_i}{\cA}{n_i}$ to the final configuration $\cA_*(\infty)= \ou{m_f}{\cA}{n_f}$. It will not be a solution to the field equations. But, it will contribute to the sum over paths in the transition amplitude $\langle m_f ,n_f,\infty  | m_i,n_i,-\infty \rangle$. We can think of this as a classically forbidden process, that nevertheless due to quantum weirdness contributes to the tunneling amplitude as a virtual process. This should all be familiar from tunneling processes in ordinary quantum mechanics. With the existence of such continuous paths connecting the initial and final configurations, all the paths have to conspire to deconstructively interfere completely to give a zero value for the transition amplitude. This is unlikely (though it could happen) so without other guiding principles we should generically expect the transition amplitude to be non-zero.

\section{Geometry from nothing: A modern approach to the Hartle-Hawking no boundary proposal \label{HH}}
I will now argue that a generalization of the instanton transition process described above can be thought of as a reformulation of the Hartle-Hawking no-boundary proposal in a more modern language of gauge theories. The proposal is different in some details, but the overall philosophy is similar, and the differences could potentially avoid some of the problems associated with the original proposal.

\subsection{The Hartle-Hawking proposal}
There are many different interpretations of the no-boundary proposal -- I will focus on the most straightforward interpretation. At its heart it is a proposal for computing the ground state of quantum gravity. The underlying assumption is that the ground state cannot be a state that transitioned from some past history -- the ground state is a state that just {\it is}. To implement this mathematically, the corresponding wavefunctional  is a path integral amplitude representing the ``transition" to a spacetime configuration from an initial state where there is no time, no space, and no geometry. This is achieved by compactifying the asymptotic past to a single point such that the manifold has no past boundary, just as the lower half of a sphere has no southern boundary (see Fig. (\ref{HH})). The proposal is then the following. The Hartle-Hawking state is a wavefunctional $\Psi[h]$, where $h$ is the three metric restricted to the present boundary $\partial M$ of a manifold $M$ with no past boundary, defined by
\beq
\Psi[h]=\int\limits^{g|_{\partial M}=h} \mathcal{D}g \ e^{i S[g_{\mu\nu}]}\,.
\eeq
It is understood in the above that for some appropriate measure, the path integral is over all metric configurations in the interior such that the pull-back of the metric to the present boundary $\partial M$ is $h_{ab}$. Diffeomorphism invariance of the gravitational action implies that the wavefunctional formally satisfies the Wheeler Dewitt equation.  

The difficulty stems from topological obstructions to filling in the interior of a space with a regular Lorentzian metric whose restriction to the boundary is a $(+,+,+)$ signature metric. Typical configurations of this sort have singularities or degeneracies at some point. To circumvent this, Hartle and Hawking supposed that there could be a period whereby the metric undergoes a phase transition from a Euclidean metric to a Lorentzian metric. The phase transition may be a genuine phase transition whereby there is some physical process that actually causes the metric signature to flip sign, or it could simply be a virtual process with no clear classical analogue corresponding to the physical reality. Nevertheless, since its inception the idea of a spacetime configuration changing metric signature has caused a great deal of confusion and controversy concerning both the physical interpretation and the practical implementation. I will now try to argue that the de Sitter gauge theory of gravity can shift the focus of the problem and thereby obviate the need for the Euclidean phase transition.

\subsection{Reformulation of the Hartle-Hawking proposal in the gauge framework of gravity}
\begin{figure}
 \begin{center}
\includegraphics[height=6cm]{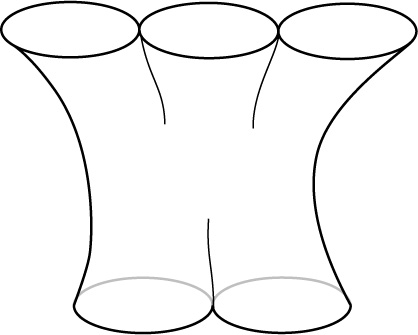}
   \end{center}
  \caption{\label{23Instanton} Here is a heuristic picture of the instanton representing a $|q|=2$ to $|q|=3$ transition. It should be understood that the actual process is a quantum tunneling process, and can't fully be illustrated by a classical geometry as above. However, this classical geometry is a typical geometry that would contribute to the path integral in the sum over paths contributing to the transition amplitude. The state asymptotically begins with the $|q|=2$ (semi-) classical solutions, and ends asymptotically with the $|q|=3$ solution.}
\end{figure}

So let's now turn to the gauge formulation of gravity to see if we can gain any insight into the nature of the proposal. Focus first on the physical interpretation of the instanton transitions in the de Sitter gauge theory. These are bonafied quantum processes, so physical intuition in the quantum regime is as difficult as it always is. However, the initial and final states have precise classical analogues. From the metric perspective, the tunneling is represented asymptotically by the transition
\begin{eqnarray}
\ou{m_f}{\bm{g}}{n_f} &\stackrel{t=+\infty}{\approx}& -dt^2+\ell^2\cosh^2(t/\ell)\,\left(q_f^2\,d\chi^2 +\sin^2q_f\chi\,\left(d\theta^2 +\sin^2{\theta} \,d\phi^2\right) \right)\nn\\
&\Big\Uparrow&  \nn\\
& & \nn\\
\ou{m_i}{\bm{g}}{n_i} &\stackrel{t=-\infty}{\approx}& -dt^2+\ell^2\cosh^2(t/\ell)\,\left(q_i^2\,d\chi^2 +\sin^2q_i\chi\,\left(d\theta^2 +\sin^2{\theta} \,d\phi^2\right) \right)
\end{eqnarray}
where as usual $q=m-n$. I have already shown how to interpret these states, so this leads to the picture in Fig. (\ref{23Instanton}).

Supposing the ground state are solutions to the quantum constraints, they annihilate the Hamiltonian constraint. In this case the transition amplitude is simply 
\beqa
\lim_{t\rightarrow \infty}\langle m_f,n_f| \mathcal{T}\exp (it\hat{\mathcal{H}}) |m_i,n_i \rangle &=& \langle m_f,n_f|m_i,n_i \rangle \nn\\
&=& \langle 0,0| \ou{m_f}{\cG}{n_f}{}^\dagger \ou{m_i}{\cG}{n_i}|0,0 \rangle \nn\,.
\eeqa
Since $\ou{m}{\cG}{n}$ forms a representation of $\pi_3(Spin(4,1))$ on the quantum Hilbert space, we should expect it to be a unitary representation. Thus, $\ou{m_f}{\cG}{n_f}{}^\dagger=\ou{-m_f}{\cG}{-n_f}$ and the transition amplitude can be written
\beq
\lim_{t\rightarrow \infty}\langle m_f,n_f| \mathcal{T}\exp (it\hat{\mathcal{H}}) |m_i,n_i \rangle =\langle \Delta m, \Delta n| 0,0\rangle.
\eeq
Thus, the important instanton transition to compute can be represented classically by the transition:
\begin{eqnarray}
\ou{\Delta m}{\bm{g}}{\Delta n} &\stackrel{t=+\infty}{\approx}& -dt^2+\ell^2\cosh^2(t/\ell)\,\left(\Delta q^2\,d\chi^2 +\sin^2\Delta q\chi\,\left(d\theta^2 +\sin^2{\theta} \,d\phi^2\right) \right)\nn\\
&\Big\Uparrow&  \nn\\
& & \nn\\
\ou{0}{\bm{g}}{0} &\stackrel{t=-\infty}{\approx}& -dt^2\,.
\end{eqnarray}

The most physically relevant transition to compute is the transition $\langle 0,1| 0,0\rangle$ representing the transition from the pre-geometric topological phase to de Sitter space:
\begin{eqnarray}
\ou{0}{\bm{g}}{1} &\stackrel{t=+\infty}{\approx}& -dt^2+\ell^2\cosh^2(t/\ell)\,\left(d\chi^2 +\sin^2\chi\,\left(d\theta^2 +\sin^2{\theta} \,d\phi^2\right) \right)\nn\\
&\Big\Uparrow&  \nn\\
& & \nn\\
\ou{0}{\bm{g}}{0} &\stackrel{t=-\infty}{\approx}& -dt^2
\end{eqnarray}
representing the birth of de Sitter space from a topological pre-geometric phase. Since our universe appears to be asymptotically approaching de Sitter space, this is the most physically relevant transition. More generically, however, one could consider the transition from the state $|0,0\rangle$ to an arbitrary configuration $\mathcal{A}_*$ in the asymptotic future. We restrict ourselves to configurations such that a polarization of the phase space $\mathcal{P}(\{\mathcal{A}_*, V_*\})$ appropriate to the canonical theory is fixed on the future boundary. The transition amplitude 
\beq
\Psi[\mathcal{P}(\{\mathcal{A}_*, V_*\})]\equiv \langle \,\mathcal{P}\{\mathcal{A}_*, V_*\}\, | 0,0 \rangle=\int\limits_{\mathcal{P}(\{\ou{0}{\mathcal{A}}{0}, V_*\})}^{\mathcal{P}(\{\mathcal{A}_*, V_*\})} \mathcal{D}\mathcal{A}\,\mathcal{D}V \ e^{i S[\mathcal{A},V]}
\eeq
can then be viewed as a wavefunctional of the configuration space variables defined on the future boundary.

\begin{figure}
 \begin{center}
\includegraphics[height=6cm]{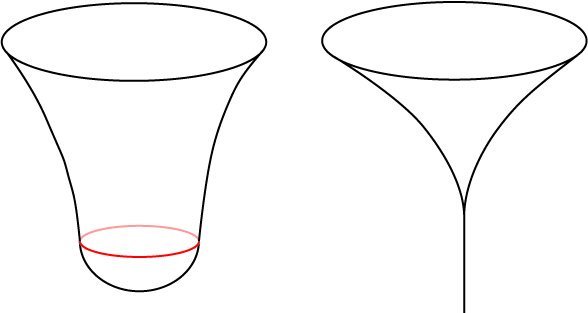}
   \end{center}
  \caption{\label{HH} On the left is the classic image representing the Hartle-Hawking proposal. This is a typical geometry representing the transition from ``nothing" to ``geometry". To match the geometry to the topology without introducing singular points, there must be a region where the geometry undergoes a phase transition (shown in red) from a Euclidean to a Lorentzian signature metric. This allows for the boundary to be capped off. On the right is our new proposal for such a transition. As before the process is a purely quantum mechanical tunneling process, but asymptotically the geometry is a topological phase with zero volume in the past. Note that although the zero volume phase is pictured as a single point, the three-space is still topologically a three-sphere.}
\end{figure}

Now compare this to the Hartle-Hawking no boundary procedure (see Fig. (\ref{HH})). The problem with the no boundary proposal can be summarized as follows. The no boundary proposal proposes to eliminate the past boundary in the path integral transition amplitude by cutting off the past end of the manifold along a three-sphere edge and gluing it back together with the closed ball $\mathbb{B}^4$ replacing the former past history. The restriction is that the pull-back of the metric to the boundary of the ball must be an ordinary three-metric with signature $(+,+,+)$. If you want to fill in the interior of the ball with a Lorentzian signature 4-metric, then you are faced with topological restrictions. Typical configurations satisfying these requirements have some pathology like a singularity or a point where the metric becomes degenerate. This is the reason why Hartle and Hawking resorted to a Euclidean phase transition -- you can fill up the closed ball with a non-pathological Euclidean 4-metric such that the restriction of the metric to the three-sphere boundary is an ordinary $(+,+,+)$ three-metric. Our proposal is different. Instead we allow for the existence of a past boundary with topology $\mathbb{S}^3$, but we subject this boundary to the severe restriction that the metric on this boundary is very special, being the configuration where the spatial 3-metric is completely degenerate. In this sense, the past boundary is still topologically and geometrically {\it contractable} to a single point, and the proposal still retains most of the ``no-boundary" flavor. Thus, the overall philosophy of the two proposals are similar -- they are both {\it geometry-from-nothing} proposals. However, this new proposal obviates the need for a hypothetical phase transition from a Euclidean to a Lorentzian metric. The gauge formulation of gravity avoids this because seemingly pathological configurations, like the metric $\ou{0}{\bm{g}}{0}$, are in fact very natural in the gauge theory of gravity. Morever, the path-connectedness of the space of connections {\it requires} that these points be taken into consideration, and that there will be non-trivial contributions to sum over paths defining the transition amplitude $\Psi[\mathcal{P}(\{\mathcal{A}_*, V_*\})]$ even without resorting to exotic contours in the complexification of the space of metrics as is often discussed in the literature of the Hartle-Hawking state. In addition to this, it places the Hartle-Hawking wavefunction on more familiar ground since the transition amplitude is directly analogous to the well understood instanton processes of non-abelian gauge theories.

\subsection{The $\theta$-vacuum \label{Theta}}
Up to this point I have used carefully chosen language to sidestep the question of whether the states $|m,n \rangle$ are true vacua. The true quantum mechanical vacuum should be stable. On the other hand, I have given hopefully convincing arguments that we should not expect the states $|m,n \rangle$ to be stable since the instanton tunneling process will cause one ``ground" state to bleed into the others. A simple way to conceptualize this is to imagine the phase space of the theory as a two-dimensional plane divided up into a grid with each square representing a $\{m,n\}$ sector. The state $|m_i, n_i \rangle$ can be thought of as a solution localized at one distinguished point in the phase space in the $\{m_i,n_i\}$ sector. In quantum mechanics, since an eigenstate of position is not an eigenstate of energy, the wave packet of a localized particle will spread rapidly. Likewise, although the concept of energy is unclear in this context, the state $|m_i,n_i\rangle$ will spread due to quantum mechanical tunneling, which is another way of rephrasing the results of the previous section. When the localized quantum state $|m_i,n_i \rangle$ relaxes, what state does it relax to? Said another way, what are the ground states of the theory that are stable against quantum mechanical tunneling? 

Stable states are easy to construct as they are familiar from Yang-Mills theories. The idea is to construct a coherent superposition of the $|m,n\rangle$ states. The state should be an equally weighted superposition of all the $|m,n\rangle$ states to eliminate tunneling. The most general superposition with this property is given by
\beq
|\Psi \rangle= \sum^{\infty}_{m'=-\infty} \sum^\infty_{n'=-\infty} e^{i\theta_1(m')}e^{i\theta_2(n')}|m',n'\rangle \,.
\eeq
Without loss of generality, we can set $\theta_1(0)=\theta_2(0)=0$. Since at this stage none of the states $|m,n\rangle$ is preferred over any other, the state $|\Psi \rangle$ should be invariant under translations of the origin generated by $\ou{\Delta m}{\cG}{\Delta n}{}^\dagger$. Since the quantum mechanical states are rays in the Hilbert space, it is sufficient to assume that these operators act projectively meaning $|\Psi \rangle $ differs from $\ou{\Delta m}{\cG}{\Delta n}{}^\dagger|\Psi \rangle $ by at most an overall phase. Putting this all together, we have
\beqa
\ou{\Delta m}{\cG}{\Delta n}{}^\dagger|\Psi \rangle &=& \ou{\Delta m}{\cG}{\Delta n}{}^\dagger\sum^{\infty}_{m'=-\infty} \sum^\infty_{n'=-\infty} e^{i\theta_1(m')}e^{i\theta_2(n')}|m',n'\rangle  \nn\\
&=& \sum^{\infty}_{m'=-\infty} \sum^\infty_{n'=-\infty} e^{i\theta_1(m')}e^{i\theta_2(n')}|m'-\Delta m,n'-\Delta n\rangle \nn\\
&=& \sum^{\infty}_{m'=-\infty} \sum^\infty_{n'=-\infty} e^{i\theta_1(m'+\Delta m)}e^{i\theta_2(n'+\Delta n)} |m',n'\rangle \nn\\
 &=& \sum^{\infty}_{m'=-\infty} \sum^\infty_{n'=-\infty} \left( e^{i(\theta_1(m'+\Delta m)-\theta_1(m'))}e^{i(\theta_2(n'+\Delta n)-\theta_2(n'))} \right) e^{i\theta_1(m')}e^{i\theta_2(n')} |m',n'\rangle \nn\\\,.
\eeqa
In order for $\ou{\Delta m}{\cG}{\Delta n}{}^\dagger$ to act projectively, the object on parentheses must be a constant phase equal to $e^{i\theta_1 \Delta m} e^{i\theta_2 \Delta n}$ where I have called $\theta_{1,2}=\theta_{1,2}(1)$. For this reason, we can label the state $|\Psi \rangle$ and just call it $|\theta_1,\theta_2 \rangle$. Like a coherent state, the state is an eigenstate of something like an annihilation operator:
\beq
\ou{1}{\cG}{0}{}^\dagger|\theta_1,\theta_2 \rangle =e^{i\theta_1}  |\theta_1,\theta_2 \rangle \quad \quad \ou{0}{\cG}{1}{}^\dagger|\theta_1,\theta_2 \rangle =e^{i\theta_2}  |\theta_1,\theta_2 \rangle \,.
\eeq

These states, known as $\theta$-states \cite{Rajaraman:Instantons}\cite{Nair:QFT} also play an important role in Yang-Mills theories and the gauge theories of the standard model. The most important conclusion to be drawn from this is that in the gauge formulation of gravity, de Sitter space (or any of the other $|m,n\rangle $ states) should not be expected to be quantum mechanically stable. To get a stable state, one has to introduce two new parameters into the theory and construct an equally weighted superposition over all the $\{m,n\}$ sectors. 

These two new continuous parameters $\{\theta_1,\theta_2\}$, which are essentially dual to the discrete parameters $\{m,n\}$, can be expected to play a role in the quantum theory of gravity and its phenomenology. Here's one simple way to see this. Consider the evolution of the state $|\theta_1,\theta_2\rangle$ from $t=-\infty$ to $t=+\infty$. One can compute this using path integrals:
\beqa
\langle \theta_1,\theta_2, \infty |\theta_1,\theta_2,-\infty \rangle &=&\lim_{t\rightarrow \infty} \sum_{\{m_i,n_i\}}\sum_{\{m_f,n_f\}} \langle m_f,n_f | e^{-i\theta_1 m_f} e^{-i \theta_2 n_f}\, \mathcal{T}\exp(i \mathcal{H} t) \,e^{i\theta_1 m_i} e^{i \theta_2 n_i} |m_i,n_i \rangle \nn\\
&=& 
=\sum_{\{\Delta m,\Delta n\}}\sum_{\{m_i,n_i\}} e^{i \theta_1 \Delta m} e^{i \theta_2 \Delta n} 
\langle m_i+\Delta m,n_i+\Delta n | \mathcal{T}\exp(i \mathcal{H} t)|m_i,n_i \rangle \nn\\
&=& \sum_{\{\Delta m,\Delta n\}} \sum_{\{m_i,n_i\}} e^{i \theta_1 \Delta m} e^{i \theta_2 \Delta n} \int^{\{m_i+\Delta m,n_i+\Delta n\}}_{\{m_i,n_i\}} \mathcal{D}\cA \mathcal{D}V  e^{i S[\cA,V]}\,.
\eeqa
In the last line, the path integral has been restricted to paths that connect the connection configuration $\{\ou{m_i}{\cA}{n_i}, V_*\}$ in the asymptotic past to $\{\ou{m_f}{\cA}{n_f}, V_*\}$ in the asymptotic future with $\{m_f,n_f\}=\{m_i+\Delta m, n_i+\Delta n\}$. 

So far nothing is new. The new stuff comes from the recognition that we can sometimes identify the instanton numbers with four-dimensional topological integrals on the manifold. To see this, it is easiest to shift from $\{\theta_1,\theta_2\}$ as dual to $\{m,n\}$ to $\{\theta ,\wt{\theta}\}$ as dual to $\{p, q\}$ where $\theta=\frac{1}{2}(\theta_1+\theta_2)$, $\wt{\theta}=\frac{1}{2}(\theta_1-\theta_2)$, $p=m+n$ and $q=m-n$. Now the integral above becomes
\beqa
\langle \theta_1,\theta_2, \infty |\theta_1,\theta_2,-\infty \rangle=\sum_{\{\Delta m,\Delta n\}} \sum_{\{p_i,q_i\}} e^{i \theta \Delta p} e^{i \wt{\theta} \Delta q} \int^{\{p_i+\Delta p,q_i+\Delta q\}}_{\{p_i,q_i\}} \mathcal{D}\cA \mathcal{D}V  e^{i S[\cA,V]}\,.
\eeqa
Suppose we could find four-dimensional functionals $\mathcal{F}_p[\cA,V]$ and $\mathcal{F}_q[\cA,V]$ such that when they are restricted to the asymptotically flat solutions $\{\ou{m}{\cA}{n}, V_*\}$ they give the relative winding numbers $\Delta p$ and $\Delta q$. Then we could rewrite the integral in terms of an effective action by
\beq
\langle \theta_1,\theta_2, \infty |\theta_1,\theta_2,-\infty \rangle=\int_{\text{All sectors}} \mathcal{D}\cA \mathcal{D}V  e^{i (S[\cA,V]+\theta \,\mathcal{F}_p[\cA,V]+\wt{\theta} \,\mathcal{F}_q[\cA,V])}
\eeq
where it is understood that the integral involves all paths connecting any asymptotically flat solution to any other. In this case the effective action, which would also be the effective action for any field theory built on this vacuum state, is $S_{effective}=S[\cA,V]+\theta \,\mathcal{F}_p[\cA,V]+\wt{\theta} \,\mathcal{F}_q[\cA,V]$. The parameters $\theta$ and $\wt{\theta}$ could be determined experimentally. 

In fact, we know one of these functionals. The second Chern class does the job:
\beqa
\mathcal{F}_p[\cA,V]=\frac{1}{8\pi^2}\int F_{\cA} \w F_{\cA}\,.
\eeqa
To see this, recall that the second Chern class is related to the Chern-Simons functional on the boundary by
\beq
\frac{1}{8\pi^2}\int_{M} F_{\cA} \w F_{\cA} =Y[\cA(\infty)]-Y[\cA(-\infty)]\,.
\eeq
This integral can be evaluated when restricted to connections that are asymptotically flat in the past and future to give $\Delta p$. Thus the parameter $\theta$ can be thought of as the coupling constant of the second Chern class of the de Sitter connection in the effective action. It has been argued in \cite{RandonoMercuri} that this parameter is related to the Immirzi parameter of Loop Quantum Gravity. Though the parameter $q$ is related to the three-volume of the flat states, as of yet I know of no de Sitter invariant four-dimensional functional that will automatically give $\Delta q$. So it is an open question as to what the parameter $\wt{\theta}$ is a coupling constant of.

\section{Summary and open problems}
Hopefully I have convinced you by now that the gauge approach to gravity is more than just empty formalism, but provides a new perspective on the nature of gravity while offering routes to potentially new physics. Here's an outline of what I discussed:
\begin{itemize}
\item{The Einstein-Cartan formalism provides the the first step to viewing gravity as a gauge theory by relegating the metric and promoting the tetrad, and introducing the spin connection.}
\item{The spin connection and tetrad can be repackaged into a single connection taking values in the dS, AdS, or Poincar\'{e} group.}
\item{The gravitational action can be written in a form that looks strikingly similar to the Yang-Mills action using these variables, \`{a} la Macdowell and Mansouri.}
\item{Recognizing that the Macdowell-Mansouri action breaks the symmetry, a new symmetry breaking mechanism can be introduced by adding new fields analogous to the Higgs.}
\item{All this can be done for any of the three groups, but the de Sitter group has the richest topological structure.}
\item{One of the unique features of gravity as a gauge theory that distinguishes it from gravity as a metric theory is the existence of topological phases in the space of solutions.}
\item{Drawing an analogy with Yang-Mills theories, the space of connection splits into ``n"-sectors, and an infinite class of new solutions to the Einstein-Cartan field equations emerge by exploiting the topological properties of the de Sitter group.}
\item{In the quantum theory, the semi-classical states associated with the ``n"-sectors form an infinite set of degenerate, quasi-stable vacua.}
\item{The path-connectedness of the space of connections likely allows for quantum mechanical tunneling between these degenerate ground states.}
\item{The transition of the state of the universe from the $|0,0\rangle$ state to an arbitrary configuration can be viewed as a twist on the Hartle-Hawking no-boundary proposal for generating a universe out of (almost) nothing.}
\item{The true stable quantum mechanical vacuum should be expected to be a coherent superposition of all the ``n"-states.}
\end{itemize}

There are a number of prominent open problems that need to be addressed. I will list some of them below:
\begin{itemize}
\item{Come up with a more physically realistic mechanism for breaking the symmetry that is fully dynamic, where some field naturally tends toward a preferred state where it takes values in $G/H$ thereby spontaneously breaking the symmetry.}
\item{Learn how to couple matter to the gauge framework of gravity in a simple and consistent way.}
\item{Determine how the degenerate surfaces affect the propagation of matter or gravitational waves.}
\item{Make sense of Quantum Field Theory on the topological background $\ou{0}{\bm{g}}{0}$.}
\item{Compute the instanton amplitude $\langle m_f,n_f| m_i, n_i \rangle$, either by semi-classical means, phase space reduction, or contour integration in the complex plane where the action is analytically extended.}
\item{Give a semi-classical approximation to the Hartle-Hawking state in the gauge gravity framework.}
\item{Determine the physical implications of the two theta-parameters of the theory.}
\end{itemize}

\section*{Acknowledgments}
Research for this article was conducted at the Perimeter Institute supported by the NSF International Research Fellowship grant OISE0853116.

\bibliography{GGBib}

\end{document}